\documentclass[preprints,article,accept,moreauthors,pdftex]{Definitions/mdpi} 
\usepackage{qtree}
\firstpage{1} 
\pubvolume{xx}
\issuenum{1}
\articlenumber{5}
\pubyear{2020}
\copyrightyear{2020}
\history{Received: date; Accepted: date; Published: date}

\Title{Transdisciplinary AI Observatory -- Retrospective Analyses and Future-Oriented Contradistinctions}

\Author{Nadisha-Marie Aliman $^{1}$, Leon Kester $^{2}$ and Roman Yampolskiy $^{3}$*}

\AuthorNames{Firstname Lastname, Firstname Lastname and Firstname Lastname}

\address{%
$^{1}$ \quad Utrecht University; nadishamarie.aliman@gmail.com\\
$^{2}$ \quad TNO Netherlands; leon.kester@gmail.com\\
$^{3}$ \quad University of Louisville;  roman.yampolskiy@louisville.edu}

\corres{Correspondence: nadishamarie.aliman@gmail.com}

\abstract{
	In the last years, AI safety gained international recognition in the light of heterogeneous safety-critical and ethical issues that risk overshadowing the broad beneficial impacts of AI. In this context, the implementation of AI observatory endeavors represents one key research direction. This paper motivates the need for an inherently \textit{transdisciplinary} AI observatory approach integrating diverse retrospective and counterfactual views. We delineate aims and limitations while providing hands-on-advice utilizing \textit{concrete practical examples}. Distinguishing between unintentionally and intentionally triggered AI risks with diverse socio-psycho-technological impacts, we exemplify a \textit{retrospective descriptive analysis} followed by a\textit{ retrospective counterfactual risk analysis}. Building on these AI observatory tools, we present near-term transdisciplinary guidelines for AI safety. As further contribution, we discuss \textit{differentiated} and tailored long-term directions through the lens of two disparate modern AI safety paradigms. For simplicity, we refer to these two different paradigms with the terms\textit{ artificial stupidity} (AS) and \textit{eternal creativity} (EC) respectively. While both AS and EC acknowledge the need for a hybrid cognitive-affective approach to AI safety and overlap with regard to many short-term considerations, they differ fundamentally in the nature of multiple envisaged long-term solution patterns.  By compiling relevant underlying \textit{contradistinctions}, we aim to provide \textit{future-oriented} incentives for constructive dialectics in practical and theoretical AI safety research. }

\keyword{AI Safety; AI Observatory; Retrospective Counterfactual Risk Analysis; Artificial Stupidity; Artificial Creativity Augmentation; Cybersecurity; Social Psychology; HCI}  

\begin{document}



\section{Motivation}

Lately, the importance of addressing AI safety, AI ethics and AI governance issues has been acknowledged at an international level across diverse AI research subfields~\cite{amodei2016concrete,dafoe2018ai,everitt2018agi,fjeld2020principled,irving2018ai,turchin2019global}. From the heterogeneous and steadily growing set of proposed solutions and guidelines to tackle these challenges, one can extract an important recent motif, namely the concept of an \textit{AI observatory} for regulatory and feedback purposes. Notable early practical realizations with diverse focuses include Italian~\cite{italianAI}, Czech~\cite{krausova2020czech}, German~\cite{germanAI} and OECD-level~\cite{oecdAI} AI observatory endeavors. Thereby, the Italian AI observatory project targets the public reception of AI technology and the Czech one tackles legal, ethical and regulatory aspects within a participatory and collective framework. The German AI observatory jointly covers technological foresight, administration-related issues, sociotechnical elements and social debates at a supranational and international level. Finally, the OECD AI Policy Observatory \textit{``aims to help policymakers implement the AI Principles"}~\cite{oecdAI} that have been pre-determined by the OECD and pertain among others to data use and analytical tools. Theoretical and practical recommendations to integrate the retrospective documentation of internationally occurring AI failures have been presented by Yampolskiy~\cite{yampolskiy2019predicting} and very recently McGregor~\cite{mcgregor2020preventing}. In addition, Aliman~\cite{aliman2020hybrid} proposed to complement such \textit{reactive} AI observatory documentation efforts with \textit{transdisciplinary} and \textit{taxonomy-based} tools as well as \textit{proactive} security activities. In this paper, we build on the approaches of both Yampolskiy and Aliman  and elaborate on the necessity of a \textit{transdisciplinary AI observatory} integrating both reactive and proactive retrospective analyses. As reactive analysis, we propose a taxonomy-based \textit{retrospective descriptive analysis} (RDA) which analytically documents factually already instantiated AI risks. As proactive analysis, we propose a taxonomy-based so-called \textit{retrospective counterfactual risk analysis}~\cite{woo2019downward} (RCRA) that inspects plausible peak \textit{downward counterfactuals}~\cite{roese2017functional} of those instantiated AI risks to craft future policies. Downward counterfactuals pertain to \textit{worse} risk instantiations that \textit{could} have plausibly happened in that specific context \textit{but did not}. While an RDA can represent a suitable tool for a qualitative overview of the current AI safety landscape revealing multiple issues to be addressed in the immediate near-term, an RCRA can supplement an RDA by adding breadth, depth and context-sensitivity to these insights with the potential to improve the efficiency of future-oriented regulatory strategies. \par

The remainder of the paper is organized as follows. In the next Section~\ref{tax}, we first introduce a simple fit-for-purpose AI risk taxonomy as basis for classification within RDAs and RCRAs for AI observatory projects. In Section~\ref{da} and in the subsequent Section~\ref{cra}, we elaborate on aims but also limitations of RDA and RCRA while collating \textit{concrete examples} from practice to clarify the proposed descriptive and counterfactual analyses. In Section~\ref{dis}, we exemplify the requirement for \textit{transdisciplinarily} conceived \textit{hybrid cognitive-affective} AI observatory approaches and more generally AI safety frameworks. In Subsection~\ref{nearterm}, we provide \textit{near-term} guidelines directly linked to the practical factuals and counterfactuals from RDA and RCRA respectively. Hereinafter, we discuss \textit{differentiated} and bifurcated \textit{long-term} directions through the lens of two recent AI safety paradigms: artificial stupidity (AS) and eternal creativity (EC) -- succinct concepts which are introduced in Subsection~\ref{longterm}. We provide incentives for future constructive dialectics by delineating central distinctive themes in AS and EC which (while overlapping with regard to multiple near-term views) exhibit pertinent differences with respect to long-term AI safety strategies. Thereafter, in Section~\ref{mat}, we briefly comment on data collection methods for RDAs and idea generation processes for RCRAs. Finally,  in Section~\ref{conc}, we summarize the introduced ensemble of transdisciplinary and socio-psycho-technological recommendations combining retrospective analyses and future-oriented contradistinctions.\par

 \section{Simple AI Risk Taxonomy}
 \label{tax}

\begin{figure}
 	
 	\centering
 	\includegraphics[width=0.45\textwidth]{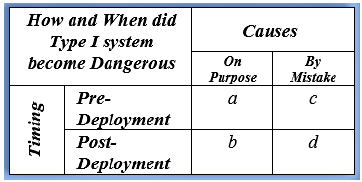}
 	\caption{Simplified overview of main Type I AI risks. Modified from~\cite{aliman2020error}.}
 	\label{dfdf}
 \end{figure}

 For simplicity and means of illustration, we utilize the streamlined AI risk taxonomy displayed in Figure~\ref{dfdf} for the classification of practical examples of AI risk instantiations in the RDA and corresponding downward counterfactuals in the RCRA. This simplified taxonomy has been derived from a recent work by Aliman et al.~\cite{aliman2020error}. (Note that the original taxonomy makes a \textit{substrate-independent} difference between two \textit{disjunct} sets of systems: \textit{Type I} systems and \textit{Type II} systems. While the set of \textit{Type II} systems includes all systems that exhibit the ability to \textit{consciously create and understand explanatory knowledge}, Type I systems are by definition all those systems that \textit{do not} exhibit this capability.  Obviously, all present-day AI systems are of \textit{Type I} whereas \textit{Type II} AI is up to now \textit{non-existent}. In fact, the only currently known sort of \textit{Type II} systems are human entities. For this reason, the taxonomy we consider here for RDA and RCRA only focuses on the practically-relevant and already instantiated classes of \textit{Type I} AI risks.) Following cybersecurity-oriented approaches to AI safety~\cite{aliman2020hybrid,brundage2018malicious,DBLP:journals/corr/PistonoY16,yampolskiy2019predicting}, we do not only classically zoom in on \textit{unintentional} failure modes but also on \textit{intentional} malice exhibited by malevolent actors. This distinction is reflected in the utilized taxonomy by contrasting AI risks brought about by malicious human actors (risk \textit{Ia} and \textit{Ib}) vs.\ those caused by unintentional failures and events (risks \textit{Ic} and \textit{Id}). Moreover, the taxonomy distinguishes between AI risks forming themselves at the pre-deplyoment stage (\textit{Ia} and \textit{Ic}) vs.\ those forming themselves at the post-deployment stage (\textit{Ib} and \textit{Id}).  
 
\section{Retrospective \textit{Descriptive} Analysis (RDA)}
\label{da}

\subsection{Aims and Limitations}
\label{a1}
 To allow for a human-centered AI governance, one requires a dynamic responsive framework that is updatable by design~\cite{Delphi2} in the light of novel emerging socio-technological~\cite{aliman2019orthogonality,cancila2018sharpening,martin2020extending} AI impacts. For this purpose, it has been postulated to combine proactive and reactive mechanisms in AI governance frameworks in order to achieve an effective socio-technological feedback-loop~\cite{Delphi2}. An RDA can be understood as a reactive AI governance and AI safety mechanism. More precisely, taxonomy-based RDA documentation efforts could facilitate a detailed especially qualitative overview and valuable opportunity for fine-grained monitoring of the AI safety landscape. It could be harnessed to guide regulatory efforts, inform policymakers and raise sensitivity in AI security, law and the general public. Further, an RDA could inform future ethical and security-aware AI design and guide endeavors to build defense mechanisms for AI systems enhancing their robustness and performance. \par In addition to the proposed fourfold qualitative distinction via the classification in risks \textit{Ia}, \textit{Ib}, \textit{Ic} and \textit{Id}, one could also introduce a quantitative parameter for intensity ratings~\cite{scott2020classification} such as harm intensity~\cite{aliman2020hybrid}. Given the harm-based nature of human cognitive templates in morality~\cite{gray2012moral,schein2018theory}, a harm parameter could provide a meaningful shortcut to encode the urgency of addressing specific risk instantiations in practice. However, given the simultaneous perceiver-dependency~\cite{gray2017think,schein2018theory} of harm perception in morality which is strongly based on dyadic considerations (the degree to which an intentional agent is \textit{perceived} to inflict damage to a vulnerable patient~\cite{schein2018theory}), corresponding assignments may not generalize. Nevertheless, identifying \textit{peaks} of harm intensity above a certain agreed upon threshold (e.g.\ starting at the level of lethal risks) from an RDA might represent a responsible strategy with less controversial assignments. (Analogously, as  further specified in Section~\ref{cra}, it is meaningful to focus on analytically derived above threshold downward counterfactuals as basis for an RCRA.) Extracted RDA peaks can be useful to calibrate regulations where necessary while avoiding superfluous constraints for multiple stakeholders that could hinder freedom and progress in the AI field.  \par Obviously, the quality of RDA results depends on data collection methods and an RDA may not reveal a comprehensive overall picture. Generally, AI risk instantiations could stay unreported, overlooked by the manual or automated data sampling or even remain unnoticed in certain contexts despite already existing. Finally, it is important to note that an RDA should \textit{not} be understood as means to \textit{predict} the future. As known from Popper, a society cannot predict the contents of its own future knowledge~\cite{popper1966poverty}. This \textit{fundamental} unpredictability is directly relevant to understand limitations of an AI observatory -- it can only reveal patterns of the past. There is no guarantee of repetitions and for instance completely unknownable novel threats could emerge via future human malevolent creativity in the form of risk instantiations \textit{Ia} and \textit{Ib} or via yet unknown errors leading to future instances \textit{Ic} and \textit{Id}. Instead of conceiving of an RDA as an oracle, we suggest framing it as a valuable preparative but incomplete tool with certain fundamental and further non-fundamental limitations. How an RCRA can be utilized to tackle one restriction of the latter sort is described in Section~\ref{cra}.\par

\subsection{RDA for AI Risk Instantiations \textit{Ia} and \textit{Ib} -- Examples}
\label{rda1}
To clarify the implementation of a taxonomy-based RDA for an AI observatory, we briefly analytically document a variety of concrete already instantiated AI risks starting with those linked to \textit{intentional} malice (AI risks \textit{Ia} and \textit{Ib}). For risk \textit{Ia}, the current goals of the human entities in the context of many induced events are mostly either \textit{adversarial} goals hold by malicious actors or \textit{research} goals of white hats and AI security researchers. To provide a simple and compact overview for risk \textit{Ia}, we group the space of these different goals in a set of 6 (unquestionably non-exhaustive) main clusters: 5 adversarial clusters and 1 research cluster conflating the research goals. The aim of the research cluster is to demonstrate the feasibility of malicious AI design motivated by diverse adversarial goals across a variety of domains in order to foster safety-awareness. Beyond that, we consider 1 extra emerging risk pattern, namely \textit{automated disconcertion}~\cite{aliman2020mal} which we introduce in a few paragraphs. \par First, an \textit{adversarial cluster 1} could be described as grouping the use of generative AI for subsequent (cyber-)crime facilitation e.g.\ via impersonation~\cite{harwell,rohrlich,rushing,stupp}. Striking examples for \textit{adversarial cluster 1} include a deep-learning based voice cloning of the CEO of a UK-based company that enabled a fraudster to acquire ca. \$243,000 ~\cite{stupp} and a scammer that suceeded to cause a transfer of ca. \$287,000 with a deepfake video sample impersonation~\cite{rohrlich}. Second, one can indentify an \textit{adversarial cluster 2} related to defamation, harassment, revenge and sextortion~\cite{gieseke2020new} typically employing deepfake techniques such as deep learning based facial replacement to visually place targeted often female individuals in pornographic video settings they never partook~\cite{ajder2019state}. Third, \textit{adversarial cluster 3} comprises the use of AI for misinformation and disinformation purposes~\cite{alba} including via fake profiles camouflaged with AI-generated synthetic portraits~\cite{reut}. Fourth, an \textit{adversarial cluster 4} consists in using deepfake methods (as well as recent applications of deepfakes to virtual reality~\cite{cole}) for a form of non-consensual voyeurism whereby even underage victims are assumed to be affected in some cases~\cite{Hao}.  Fifth, \textit{adversarial cluster 5} includes AI-supported espionage~\cite{corera} (e.g.\ via AI-generated fake profile pictures on social media platforms\cite{satter}), AI-aided intelligence gathering~\cite{probyn} and controversial AI-supported targeted profiling~\cite{mozur}. \par Moreover, we identify a \textit{research cluster 1} as described. Notably, security researchers provided proof-of-concepts among others related to designing camouflaged undetectable fake samples usable for other crimes (e.g. adversarial deepfakes bypassing deepfake filters~\cite{neekhara2020adversarial} which could be misused to conceal unethical illegal material disguised as deepfakes and furthermore undetected AI-generated fake comments i.a.\ on a federal public comment website~\cite{zangt}). Recent security work also successfully explored advanced deepfake techniques for improved impersonation, spear-phishing and large-scale disinformation~\cite{donnell}. Yampolskiy crafted a proof-of-concept for an AI-generated fake academic article~\cite{TransformerJrManuscript-GPTWSC} perhaps simultaneously acting as cautionary example and as a form of honeypot~\cite{zhang2003honeypot} for inattentive readers that might cite this article unknowingly. Other researchers identified an emerging interest for deepfake ransomware~\cite{nelson} in certain cybercriminal circles. Beyond that,  it has been demonstrated that via a replica of a victim intelligent system (a deep reinforcement learning agent), the policies of the victim system can be compromised in a targeted way~\cite{chen2019adversarial}.  \par Interestingly, an already perceptible consequence of \textit{the mere existence} of risk \textit{Ia} instantiations containing the design of deepfake technologies already led to the emergence of a risk pattern which has been termed automated disconcertion~\cite{aliman2020mal}. Automated disconcertion can imply the intentional or also unintentional mislabelling of real samples as fake -- e.g.\ in the context of misleading conspiracy theories~\cite{spocchia} or against the background of uncertain political settings as it was the case in Gabon not long ago~\cite{Hao2}. (To summarize the latter, a \textit{"recent failed
military coup in the context of pre-existing political unrest in
Gabon was partially grounded in the proliferation of the wrong
assumption that an official presidential video represented a
manipulative deepfake video"}~\cite{aliman2020mal}.) Conversely, automated disconcertion can also mean that fake samples are considered as being authentic or simply lead to highly uncertain and inconclusive settings in which doubts cannot be further resolved in reasonable time with acceptable resources. In short, this additional outlier risk pattern is called \textit{automated} disconcertion since it does not further necessitate the interference of any actors to be repeatedly instantiated after initiation.  \par 

Coming to risk \textit{Ib}, its instantiations are currently predominantly concentrated in a single research-oriented cluster (in analogy to \textit{research cluster 1} for risk \textit{Ia} instantiations). However, it is thinkable that exploits of AI vulnerabilities unknown to the public are already taking place before disclosure (a type of zero-day exploits~\cite{bilge2012before} applied to the AI domain). The main benign research goal for security researchers to target risk \textit{Ib} instantiations is currently mostly to disclose existing AI vulnerabilities against malicious attacks and explore possible novel defenses against those before their exploitation. This already led to an incessant attacker-defender race in the fast moving field of security for machine learning and adversarial examples~\cite{carlini2017adversarial,carlini2020partial,papernot2016towards,tramer2020adaptive}. In recent years, researchers have among others developed different attack schemes on how to evade cybersecurity AI~\cite{kirat2018deeplocker}, e-mail protection, verification tools~\cite{qiu2019semanticadv}, forensic classifiers~\cite{carlini2020evading} and person detectors~\cite{xu2020adversarial}, how to elicit algorithmic biases~\cite{aliman2020hybrid,wallace2019universal}, how to fool medical AI ~\cite{cheng2020adversarial,finlayson2019adversarial,han2020deep,zhang2020tiny}, law enforcement tools~\cite{zhou2018invisible} as well as autonomous vehicles~\cite{cao2019adversarial,mcafee}, how to perform denial-of-service and other adversarial attacks on commercial AI services~\cite{247642,li2019adversarial,wu2020decision}, how to cause energy-intense and unnecessarily prolonged processing time~\cite{shumailov2020sponge} and how to poison AI systems post-deployment~\cite{cina2020black}.

\subsection{RDA for AI Risk Instantiations \textit{Ic} and \textit{Id} -- Examples}
\label{rda2}
In this subsection, we continue to elucidate the practical application of a taxonomy-based RDA by now briefly analytically documenting various already instantiated \textit{unintentionally} triggered risks that formed themselves at the pre- and post-deployment stage (i.e.\ risk \textit{Ic} and \textit{Id} respectively). For risk \textit{Ic}, we group the space of observed failure modes in a set of 5 (unquestionably non-exhaustive) main failure clusters. In addition, we present 1 extra emerging risk pattern. In analogy to the outlier risk pattern of automated disconcertion related to risk \textit{Ia} instantiations, we introduce the risk pattern of \textit{automated peer pressure} representing an already perceptible side-effect of specific risk instantiations \textit{Ic}. In the case of AI risk instantiations \textit{Id}, we consider a single main failure cluster. (Overall, in some cases, it is difficult to delineate a risk instantiation type unambiguously (e.g.\ \textit{Ic} vs. \textit{Id} in the presence of multiple complex influences or even in a few cases \textit{Ic} vs. \textit{Ia} given different ethical perspectives). This practical limitation is partially linked to the perceiver-dependency of classification-related assignments that may also play a role in a future AI observatory. However, by publicly sharing the sources, it is possible for entities external to an AI observatory to refine interpretations. Generally, we humbly subscribe to the epistemological view that all knowledge is fallible~\cite{chitpin2013should}.)\par  
For risk \textit{Ic}, we consider the 5 main failure clusters described in the following. First, \textit{failure cluster 1} comprises ethically-relevant instances of algorithmic bias~\cite{obermeyer2019dissecting}. Part of this cluster are misclassifications of diverse underrepresented patterns in AI training datasets with unethical repercussions as exhibited in e.g.\  facial misidentification~\cite{hill2020wrongfully}, facial recognition failures~\cite{buolamwini2018gender,joy}, inaccuracy in AI-aided diagnosis~\cite{larrazabal2020gender}. Other cases are datasets with historically outdated unethical labels~\cite{prabhu2020large} and ethically-sensitive training biases favoring overrepresented patterns~\cite{jain2020imperfect}. Second, \textit{failure cluster 2} refers to instances of poorly designed low-performing AI that are halted subsequently~\cite{kempsell}. Third, \textit{failure cluster 3} are AI methods designed for law enforcement but threatening privacy~\cite{huchel}. Fourth, \textit{failure cluster 4} subsumes all unintentional risk instantiations linked to more or less hidden pseudo-scientific or outdated and previously refuted preconceptions. For instance, the deployment of AI for facial recognition of criminals based on \textit{"minute features"}~\cite{cushing,facial2} in their face is based on pseudo-scientific assumptions~\cite{facial3}. Further, the deployment of present-day image-based "emotion recognition" AI is not grounded in state-of-the-art~\cite{barrett2019emotional} affective science and lacks the required multimodal and context-sensitive modelling to be able to mimick how humans \textit{infer}~\cite{gendron2020emotion} (and not detect) affective patterns. In fact, a ban has been requested for premature emotion AI i.a.\ to prevent usage in ethically sensitive settings~\cite{now} such as law enforcement, fraud detection or recruiting. Fifth, \textit{failure cluster 5} is linked to affective, persuasive~\cite{lieber} and (micro-)targeted AI-aided methods that already permeate human cognitive-affective constructions in a way extending \textit{beyond the initial design purposes} and causing \textit{epistemic} biases ranging from a loss of critical stance via AI-empowered social media~\cite{jakubowski2019s,sawers} to flawed mind perception in present-day robots~\cite{chikhale2018multidimensional,yam2020robots}.\par
 A further risk pattern that emerged via \textit{the mere existence} of specific AI risk instantiations \textit{Ic} assignable to the \textit{failure cluster 5}, is a construct that we call \textit{automated peer pressure}. It is already known that attention at a collective level can be \textit{intentionally} biased and manipulated in social media~\cite{jakubowski2019s} also with the help of bots~\cite{orabi2020detection,prier2017commanding} (risk \textit{Ia}). Moreover, as stated in an open letter written by multiple known psychologists and sent to the American Psychological Association: \textit{``[...] the desire for social acceptance and the fear of social rejection are exploited by psychologists and other behavior change experts to pull users into social media sites and keep them there for long periods of time"}~\cite{apapsy} -- especially children~\cite{lieber}. Susceptible collective attention mechanisms and beliefs are already even \textit{unintentionally}~\cite{sawers} strongly influenced by AI-empowered social media  initially developed for benign purposes. Paired with the strong social dependency of humans where social pressure plays an important regulatory role with biological roots~\cite{theriault2020sense}, it already triggered what one could call automated peer pressure, a self-perpetuating pattern of social pressure~\cite{anderson2018teens,barbera2018new,franchina2018influence,halfmann2019permanently,stieger2018week} without the need for social agents that directly and consciously exert it. Beyond that, the known group phenomenon of \textit{``self-reinforcing networks of like-minded users"}~\cite{prier2017commanding} encountered in social media has been termed \textit{homophily}~\cite{jakubowski2019s,prier2017commanding}. Overall, a combination of a multiplicity of heterogeneous factors of which epistemic biases, homophily, affective contagion~\cite{ferrara2015measuring,jakubowski2019s}, bots and automated peer pressure are only a subset may foster the documented spread of propaganda in social media~\cite{prier2017commanding} as well as the reported negative impacts on the mental health of young users~\cite{luxton2012social,sawers}.  \par

   Finally, concerning AI risk \textit{Id}, we observe one main failure cluster which is connected to unanticipated post-deployment usage modes and contexts which also includes eventual complications within unusual interactions of the AI system in a dynamically changing environment. Notable examples are failures of facial recognition AI linked to COVID-19 causing the widespread use of facial masks~\cite{lane2020nist,mundial2020towards,ngan2020ongoing}, the invariant responses of natural language processing systems when faced with nonsensical instead of usual meaningful queries~\cite{krishna2019thieves} (disclosing the low level of understanding) and the AI-based censorship of a picture displaying ancient slavery settings due to a forerunning misclassification labelling the sample as displaying nudity~\cite{taylorj}. Other cases include unknown latent biases in medical AI~\cite{decamp2020latent} and other forms of biases in medical AI that unfold post-deployment as a function of geographical factors~\cite{kaushal2020geographic}.

\section{Retrospective \textit{Counterfactual} Risk Analysis (RCRA)}
\label{cra}

\subsection{Aims and Limitations} 
\label{arcra}
While \textit{upward counterfactuals} of a factual event refer to the better ways in which that event could have unfolded but did not, \textit{downward counterfactuals} refer to those conceivable ways in which this event could have turned out worse. In the past, counterfactual thinking has often been framed as detrimental rumination or even as cognitive bias. However, a modern explanatory framework from social psychology termed \textit{functional theory of counterfactual thinking} (abbreviated with FTCT in the following) stresses that counterfactual thoughts can offer \textit{``[...] insights that comprise blueprints for future action [...]"}~\cite{roese2017functional}. FTCT stresses that counterfactual thinking serves problem-solving and can exhibit high usefulness especially in \textit{complex multi-causal domains}~\cite{roese2017functional}. At the intrapersonal level, counterfactual thoughts are based on implicit processes caused by problems, they are linked to a negatively valenced state of core affect~\cite{epstude2008functional} and have the potential to evoke (mental or physical) actions that can potentially correct the underlying errors. This procedure instantiates a regulatory loop -- which corresponds to a type of negative feedback model~\cite{epstude2008functional} enacted as goal-oriented corrective behavior.\par  Recently, the notion of an RCRA~\cite{woo2019downward} building upon downward counterfactuals from historical events has been proposed to risk stakeholders in the context of risk management applied to hazardous events (such as earthquakes or terroristic attacks). As explained by Woo~\cite{woo2019downward}, such an innovative augmented historical analysis represents a generic universal tool that can supplement regulatory resilience tests and sense-making while facilitating the formation of more differentiated and nuanced views. Given its conjectured \textit{domain-general} nature and seeming applicability to complex multi-causal domains of risk analysis, we suggest to transfer RCRA to AI observatory contexts \textit{at a conceptual level}.\par  

For illustration purposes, the Subsection~\ref{rcraex} presents a simplified \textit{RDA-based RCRA} which directly builds upon the exemplary RDA performed in the last Section~\ref{da}. Our method is \textit{loosely} inspired by Woo's RCRA conception which manifests itself by the general integration of downward counterfactuals from historical samples. However, our step-wise methodology (elucidated in the subsequent Subsection~\ref{proc}) to extract meaningful candidates for the simulation\footnote{Downward counterfactuals can be (co-)created e.g.\ in a predominantly mental form, facilitated by immersive design fiction settings~\cite{aliman2020mal} (including storytelling narratives and virtual reality) or simulated and visualized with technological tools.} of downward counterfactuals given a large state space of past events has been independently conceived and tailored to the specific AI observatory domain. Overall, we understand an RCRA as complement for a forerunning RDA. Together, this pair of retrospective analyses could provide a solid starting point for future AI observatory projects to be however necessarily updated and error-corrected with time. \par In abstract terms, combining RDA and RCRA can be seen as a socio-technological enactment of the regulatory loop-governing behavior~\cite{epstude2008functional} described in FTCT -- which fits to the AI governance recommendation mentioned earlier in Subsection~\ref{a1}, namely the notion of a socio-technological feedback-loop combining proactive and reactive measures~\cite{aliman2020hybrid,aliman2019orthogonality,Delphi2}. While an RDA mainly represents a \textit{reactive} documenting approach, an RCRA \textit{attempts} to broaden future \textit{proactive} measures by anticipating potential extreme branches of the future while resisting the fallacy to cast itself as oracle tool. We emphasize that in the light of the fundamental unpredictability of future knowledge creation as well as the fallibility and incompleteness of human knowledge, surprises and errors are unavoidable. No RCRA can guarantee unassailability. This is similarly the case in cybersecurity for other types of techniques that are likewise assignable to a broad class of proactive security measures related to downward counterfactuals such as penetration testing~\cite{weidman2014penetration} and red teaming~\cite{rajendran2011blue,rege2016incorporating}. Also there it holds that the non-detection of a vulnerability does not guarantee its absence. (Conversely, the detection of a vulnerability does also not guarantee its future exploitation by malicious actors\footnote{For instance, their interest could shift, the asset could be(come) less interesting or the attack too time-consuming and costly.}.)\par

\subsection{Preparatory Procedure} 
\label{proc}

After having expounded on aims and limitations of an RDA-based RCRA, we speak to the preparatory procedure of meaningfully extracting the required downward counterfactuals for an RCRA taking as input the set $ O_{RDA} $ containing \textit{all instances} from the forerunning RDA. However, before providing further details, we recall as mentioned in Subsection~\ref{a1} that a meaningful agreed upon threshold $ \tau $ of harm intensity is recommendable. Although perceiver-dependent, a sufficiently high threshold such as e.g.\ when set to plausible downward counterfactuals of \textit{at least} \textit{lethal} dimension may be suitable.  On an oversimplified harm intensity scale with \textit{1} standing for almost no harm and \textit{5} for existential risk, let \textit{4} stand for a lethal risk (with \textit{2} encoding minor and \textit{3} major harm). Naturally, this threshold and scale are solely employed for purely illustrative purposes and more differentiated and tailored approaches may be required in practice~\cite{aliman2020hybrid}. Equipped with the scale and the exemplary threshold $ \tau=4 $, we elaborate in the following on how the set $ O_{RCRA} $ of all \textit{clusters}\footnote{For an enhanced context-sensitivity and to  avoid overfitting to the idiosyncrasies of single isolated events, we recommend RCRA simulations at the level of clusters and not of single instances as becomes apparent in the next Subsection~\ref{rcraex}.} considered in an RCRA can be constructed starting with $ O_{RDA} $ and consecutively applying the following ordered sequence of 4 operations in yet to be described ways: 1) \textit{taxonomization}, 2) \textit{analytical clustering}, 3) \textit{brute-force deliberation and threshold-based pruning}, 4) \textit{assembly}.\par

 \begin{figure}
	
	\centering
	\includegraphics[width=0.75\textwidth]{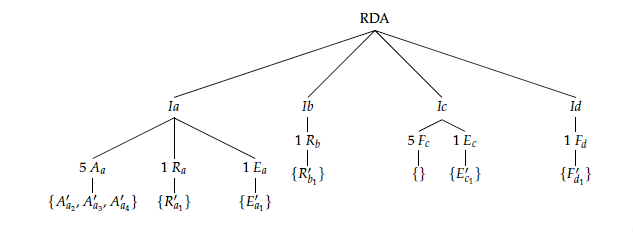}
	\caption{Simplified sketch on possible preparatory procedure to extract peak generic downward counterfactuals for an RCRA out of a forerunning taxonomy-based RDA for an AI observatory. The top node stands for the initial set $ O_{RDA} $ containing all RDA samples. For illustration, the risk instantiation clusters from Section~\ref{rda1} and~\ref{rda2} are filled in. $ A $ refers to \textit{adversarial}, $ R $ to \textit{research}, $ E $ to \textit{extra} and $ F $ to \textit{failure} cluster. The conjunction of all analytically derived leaves are possible generic above threshold downward counterfactuals of interest for the RCRA. In this example, the output set for the RCRA corresponds to $ O_{RCRA}=\{ A'_{a_2}, A'_{a_3},A'_{a_4},R'_{a_1},E'_{a_1},R'_{b_1},E'_{c_1},F'_{d_1}\}  $. For more details, see text. } 
	\label{rcratree}
\end{figure}
As first step, \textit{taxonomization} is applied to $ O_{RDA} $ which consists in a one-to-one mapping of each AI risk instantiation sample to either a key from the taxonomy (i.e.\ \textit{Ia}, \textit{Ib}, \textit{Ic} or \textit{Id}) or in theory to a generic placeholder key for novel unknown patterns. In our description, all samples were directly or at least secondarily assignable to the pre-existing taxonomy keys and no unknown key was required. As second step, the researchers apply an \textit{analytical clustering} operation based on a self-generated explanatory semantic grouping linking every sample associated with a specific key, to a cluster. By way of example, under risk \textit{Ia} discussed in Section~\ref{rda1}, this operation led to 5 adversarial clusters, 1 research cluster and 1 extra cluster. In a third step, the researchers apply \textit{brute-force deliberation\footnote{In theory, this search can be optimized further. However, the aim is to (at a later stage) obtain \textit{a broad as possible} set of counterfactual instances to increase illustrative power. Both one-to-one and many-to-one mappings between downward counterfactual instances and clusters can potentially become RCRA-relevant if stored. This is connected to the \textit{complementary cognitive co-creation} method used to interlink the preparatory procedure with the RCRA that we explain in Subsection~\ref{matrcraprep}.}  and threshold-based pruning} by mentally going through every single sample of $ O_{RDA} $ and trying to devise -- within reasonable self-determined time limits -- a plausible downward counterfactual where it holds for the self-rated harm intensity $ h $, that $ h\geq\tau $. If such a suitable downward counterfactual is generated in time, the sample is \textit{maintained}, otherwise the sample is discarded from further consideration. Finally, the fourth operation \textit{assembly} is performed which requires to assemble $ O_{RCRA} $ by linking back the remaining samples to their clusters from the second step. On this basis, one obtains the generic downward counterfactuals that need to be analyzed for the intended RCRA. In short, this simple step-wise procedure takes RDA instances as inputs and produces a set of generic RCRA clusters as output. This output set $ O_{RCRA} $ represents the superset of the searched meaningful generic above threshold downward counterfactuals of interest. For clarification, the next paragraph briefly comments on the application of this simple preparatory procedure to our exemplary RDA instances.\par

 While applying the third step of brute-force deliberation and threshold-based pruning, we deleted a large amount of RDA samples since many instances did \textit{not} seem to have had a plausibe downward counterfactual with a harm intensity $ h\geq\tau $. However, we decisively already identified certain rare samples where this condition was fulfilled. In the fourth step, we assembled $ O_{RCRA} $ by linking these maintained samples back to 8 RDA clusters as described in the following. For risk \textit{Ia}, we deleted the first and fifth adversarial cluster but maintained \textit{adversarial cluster 2} (encoded with $ A_{a_2} $), \textit{adversarial cluster 3} ($ A_{a_3} $), \textit{adversarial cluster 4} ($ A_{a_4} $), the single \textit{research cluster} ($ R_{a_1} $) and the extra cluster ($ E_{a_1} $)  of \textit{automated disconcertion}. For risk \textit{Ib}, the standalone research cluster ($ R_{b_1} $) was maintained. For risk \textit{Ic}, only the extra cluster ($ E_{c_1} $) of \textit{automated peer pressure} remained, and we deleted all failure clusters. Finally, for risk \textit{Id}, we kept the single available failure cluster ($ F_{d_1} $). While these clusters were mapped to \textit{factual }risk instantiations, an RCRA obviously requires the generation of corresponding downward counterfactuals. Thus, instead of $ A_{a_2} $, we encode its unreal generic downward counterfactual which we denote $ A'_{a_2} $. Similarly, instead of $ A_{a_3} $, we write $ A'_{a_3} $ and so forth. Consequently, as illustrated in a highly simplified form in Figure~\ref{rcratree}, one can hereafter fairly straightforwardly assemble these fragments (visualized as the leaves of the tree) in order to obtain the final output set $ O_{RCRA}=\{ A'_{a_2}, A'_{a_3},A'_{a_4},R'_{a_1},E'_{a_1},R'_{b_1},E'_{c_1},F'_{d_1}\}  $. \par

\subsection{Exemplary RDA-based RCRA for AI Observatory Projects}

\label{rcraex}

Recently, co-creation \textit{design fictions} (DFs)~\cite{ahmadpour2019co,pillai2020communicate} known from human-computer interaction (HCI)~\cite{rapp2020design} have been recommended for security practices in the AI field~\cite{houde2020business} and at the intersection between AI and virtual reality (AIVR)~\cite{aliman2020mal}. Generally, DFs \textit{``can be used for technological future projections by experts in
the form of e.g. narratives or construed prototypes that can
be represented in text, audio or video formats but also in VR
environments"}~\cite{aliman2020mal} (whereby the authors cautioned likewise \textit{not} to regard a DF as a means to \textit{predict} the future but as preparatory tool). In our view, one promising way to perform an RDA-based RCRA could be to frame each RCRA cluster as co-creation DF task. Distinctively, instead of projecting into the future as performed in classical DF contexts, such an \textit{RDA-based RCRA-DF} construes instances of RCRA cluster narratives (or experiential prototypes) projecting to the counterfactual past. For illustration, we apply a simplified RDA-based RCRA-DF to each of the 8 elements within the set $ O_{RCRA}=\{ A'_{a_2}, A'_{a_3},A'_{a_4},R'_{a_1},E'_{a_1},R'_{b_1},E'_{c_1},F'_{d_1}\}  $ assembled in the preparatory procedure of the previous Subsection~\ref{proc}. For RCRA-DFs pertaining to intentional malice (risks \textit{Ia} and \textit{Ib}), we provide short DF narratives taking the form of a succinct \textit{threat model}~\cite{aliman2020mal,carlini2019evaluating} specifying adversarial goals, knowledge and capabilities. By contrast, for unintentional failure modes (risks \textit{Ic} and \textit{Id}), we instead describe a short failure model comprising initial design goals, knowledge gaps and unintended effects. Generally, we only consider instantiations of RCRA clusters that correspond to above threshold downward counterfactuals (i.e.\ with a harm intensity $ h\geq \tau $ whereby in Subsection~\ref{proc}, $ \tau $ was exemplarily set to lethal risk dimensions).

 \subsubsection{Downward Counterfactual DF Narrative $A'_{a_2}$ }
 \label{aa2}
\begin{itemize}
	\item \textbf{\textit{Adversarial Goals:}}  AI-aided defamation, revenge, harassment and sextortion.
		\item \textbf{\textit{Adversarial Knowledge:}} Since it is a malicious stakeholder that is designing the AI, the system is available to this adversary in a transparent\textit{ white-box} setting. Concerning the knowledge pertaining to the human target, a \textit{grey-box} setting is assumed. Open-source intelligence gathering and social engineering are exemplary tools that the adversary can employ to widen its knowledge of beliefs, preferences and personal traits exhibited by the victim.
			\item \textbf{\textit{Adversarial Capabilities:}} In the following, we briefly speak to exemplary plausible counterfactuals of at least lethal nature that malicious actors could have been capable to bring about and that are \textit{``worse than what actually happened"}~\cite{woo2019downward} (as per RDA). For defamation purposes, it would have been for instance possible to craft AI-generated fake samples that wrongly incriminate victims with \textit{not} actually executed actions (e.g.\ a fake homicide but also fake police violence) leading to a subsequent assassination when deployed in precarious milieus with high criminality. To enact revenge with lethal consequences in socio-cultural settings that particularly penalize the violation of restrictive moral principles, similar AI-based methods could have been applicable (e.g.\ via deepfakes assumingly displaying fake adultery or contents linked to homosexuality). An already instantiated form of AI-enabled  harassment mentioned in the RDA consists in sharing fake AI-generated video samples of pornographic nature via social media channels~\cite{ajder2019state}. Consequences could include suicide of vulnerable targets (as generally in cybervictimization~\cite{john2018self}) or exposure to a lynch mob. In fact, the contemplation of suicide by deepfake pornography targets has already been reported lately~\cite{gieseke2020new}. Finally, concerning AI-supported sextortion, warnings directed to teenagers and pertaining to the convergence of deepfakes and sextortion have been formulated recently~\cite{crothers}. Given the link between sextortion and suicide associated with motifs such as i.a.\ hopelessness, humiliation and shame~\cite{nilsson2019understanding}, consequences of technically feasible but not yet instantiated deepfake sextortion scams could also include suicide -- next to simplifying this criminal enactment by adding automatable elements.

\end{itemize}

 \subsubsection{Downward Counterfactual DF Narrative $A'_{a_3}$ }
 \label{aa3}
 \begin{itemize}
 	\item \textbf{\textit{Adversarial Goals:}} AI-aided misinformation and disinformation.
 \item \textbf{\textit{Adversarial Knowledge:}} Identical to adversarial knowledge indicated in~\ref{aa2}.
 \item \textbf{\textit{Adversarial Capabilities:}} Technically speaking, a malicious actor could have crafted misleading and disconcerting fake AI-generated material that could be interpreted by extreme endorsers of pre-existing misguided conspiracy theories as providing evidence for their beliefs inciting them to subsequent lethal violence. A historical precedent of gun violence as reaction to fake news seemingly confirming false conspiracy theories was the Pizzagate shooting case where a young man fired a rifle in a pizzeria \textit{"[...] wrongly believing he was saving children trapped in a sex-slave ring"}~\cite{haag}. Beyond that, when it comes to (micro-)targeted~\cite{jakubowski2019s} disinformation, conceivable malicious actors could have more systematically already employed hazardous AI-aided information warfare~\cite{prier2017commanding} techniques in social media. This could have been supported by AI-enabled psychographic targeting tools~\cite{jakubowski2019s} and via networks of automated bots~\cite{bessi2016social,prier2017commanding} partially concealed via AI-generated artefacts such as fake profile pictures. While the level of sophistication of many present-day social bots is limited~\cite{assenmacher2020demystifying}, more sophisticated bots emulating a breadth of human online behavior patterns are already developed~\cite{boneh2019relevant,yang2019arming} and it is known for some time~\cite{shao2018spread} that \textit{``[...] political bots exacerbate political polarization"}~\cite{yan2020asymmetrical}. By AI-aided microtargeting of specific groups of people that are ready to carry out violent acts, malicious actors could have caused more political unrest with major lethal outcomes. In fact, Tim Kendall who was a prior director of monetization at Facebook recently stated more broadly that \textit{``[...] one possible near-term effect of online platforms' manipulative and polarizing nature could be civil war"}~\cite{sawers}.

\end{itemize}

  \subsubsection{Downward Counterfactual DF Narrative $A'_{a_4}$ }
  
\begin{itemize}
	\item \textbf{\textit{Adversarial Goals:}} AI-aided non-consensual voyeurism.
	\item \textbf{\textit{Adversarial Knowledge:}} Identical to adversarial knowledge indicated in~\ref{aa2}.
	\item \textbf{\textit{Adversarial Capabilities:}} Before delving into downward counterfactuals that corresponding malicious actors could have already brought about, it is important to note that the goal considered in this cluster is not primarily the credibility or appearance of authenticity exhibited by the synthetic AI-generated contents. Rather, the focus when visually displaying the target non-consensually in compromising settings is more on feeding personal fantasies or facilitating a demonstration of power~\cite{cole,Hao} while the synthetic samples can obviously concurrently be shared via social media channels. Against this backdrop, it is not difficult to imagine that when editing visual material of vulnerable targets with practices such as deep-learning based "undressing"~\cite{Hao}, a disclosure could induce motifs of hopelessness, humiliation and shame in some of those individuals provoking suicidal attempts similar to the hypothetical deepfake sextortion counterfactual described in~\ref{aa2}. The mere sensing of having been victimized via non-consensual deepfake pornography has also been associated with the perception of a "digital rape"~\cite{farokhmanesh2018legal,gieseke2020new}. Especially when the victims are underage~\cite{Hao}, this could plausibly reinforce suicidal ideation. Another dangerous avenue may be subtle \textit{combination possibilities} available to the malicious actor. Non-consensual voyeuristic (but also more generally abusive) illegal but quasi-untraceable material bypassing content filters could be meticulously concealed with deepfake technologies and unnoticedly propagated\footnote{For instance by mixing real material with synthetic elements obtained from style-based generative adversarial network methods~\cite{karras2020analyzing}, deep-learning based face-replacement and \textit{adversarial deepfake} techniques~\cite{neekhara2020adversarial} in order to evade content filters critical to law enforcement.} for some time. This could hinder criminal prosecution and particularly threaten the life of vulnerable young victims. Potentiated with automated disconcertion, it could cause a set of latent lethal socio-psycho-technological risks.
\end{itemize}
 
  \subsubsection{Downward Counterfactual DF Narrative $R'_{a_1}$ }
  \label{ra1}
   
\begin{itemize}
	\item \textbf{\textit{Adversarial Goals:}} Research on malevolent AI.
	\item \textbf{\textit{Adversarial Knowledge:}} Identical to adversarial knowledge indicated in~\ref{aa2}.
	\item \textbf{\textit{Adversarial Capabilities:}} To begin with, note that in this RCRA cluster, we assume that the research is motivated by \textit{malign} intentions contrary to the corresponding factual RDA research cluster that is conducted with benign and precautionary intentions by security researchers and white hats as mentioned in Subsection~\ref{rda1}. This additional distinction is permissible due to its property as \textit{downward} counterfactual. By way of illustration, malicious actors could have already performed research on malevolent AI design in the domain of autonomous mobility or in the military domain. They could have developed a novel type of meta-level physical adversarial attacks on intelligent systems\footnote{With intelligent systems, we refer to technically feasible AIs able to independently perform the OODA-loop (i.e.\ observe, orient, decide, act), but simultaneously totally subordinated to and goal-governed by human entities (e.g.\ using updatable \textit{human-defined} ethical goal functions~\cite{Delphi2} prepared pre-deployment -- where humans fill in ethically-relevant parameters into a suitable blank but context-sensitive scientifically-grounded template denoted augmented utilitarianism~\cite{aliman2020hybrid}).} directly utilizing other physically deployed intelligent systems under their control. Such an attacker-controlled intelligent system could be employed as a new advanced form of present-day physical adversarial examples~\cite{chen2020devil,duan2020adversarial,kong2020physgan,nassi2020phantom,wang2020towards,xu2020adversarial} against a selected victim intelligent system. The maliciously crafted AI could have been designed to optimize on physically fooling the deployed victim system e.g.\ via physical manipulations at the sensor level such as to misleadingly bring about victim policies with lethal consequences entirely unintended by the operators of the victim model. A further concerning instance of malign research could have been secretive or closed-source research on automated medical AI forgery tools that add imperceptible adversarial perturbations to inputs such as to cause tailored customizable misclassifications. While the vulnerability of medical AI to adversarial attacks is already known~\cite{cheng2020adversarial,finlayson2019adversarial,han2020deep,rahman2020adversarial,zhang2020tiny}  and could be exploited by actors intending medical fraud e.g.\ for financial gains, certain exertions of this practice in the wrong settings could be misused as tool for murder attempts and targeted homicides.

\end{itemize}
 
    \subsubsection{Downward Counterfactual DF Narrative $E'_{a_1}$ }
    \label{ea1}
    
    \begin{itemize}
    	\item \textbf{\textit{Adversarial Goals:}} This extra cluster of automated disconcertion refers to a risk pattern that emerged automatically from the mere availability and proliferation of deepfake methods in recent years. However, it is conceivable that this AI-related agentless automatic pattern can be intentionally instrumentalized in the service of other (not necessarily AI-related) primary adversarial goals. One example for a primary adversarial goal cluster in the light of which it is appealing for a malicious actor to strategically harness automated disconcertion, would be \textit{information warfare and agitation on social media}. In fact, early cases may already occur~\cite{spocchia}. 
  
    	\item \textbf{\textit{Adversarial Knowledge:}} Identical to adversarial knowledge indicated in~\ref{aa2}.
    	\item \textbf{\textit{Adversarial Capabilities:}} The use of social media in information warfare has been described to be linked to the objective to intentionally blur the lines between fact and fiction~\cite{jakubowski2019s}. The motif of automated disconcertion itself could be weaponized and misleadingly framed as providing evidence for post-truth narratives offering an ideal breeding ground for global political adversaries performing information warfare via disinformation. Malicious actors could then intensify this framing with the use of pertinent AI technology enlarging their adversarial capabilities as described earlier under the cluster of AI-aided misinformation and disinformation in~\ref{aa3}. Given that automated disconcertion may aggravate pre-existing global strategically maintained confusions~\cite{ciosek2020aggravating}, it becomes clear that a more effective incitement to lethal violence, political unrest with major lethal outcomes or civil wars could be achieved.

    \end{itemize}
   \subsubsection{Downward Counterfactual DF Narrative $R'_{b_1}$ }   
\begin{itemize}
	\item \textbf{\textit{Adversarial Goals:}} Research on vulnerabilities of deployed AI systems.
	\item \textbf{\textit{Adversarial Knowledge:}} \textit{Grey-box} setting (partial knowledge of AI implementation details).
	\item \textbf{\textit{Adversarial Capabilities:}} As analogously described in~\ref{ra1}, we assume that the research is conducted with malicious intentions. Zero-day exploits of vulnerabilities in (semi-)autonomous mobility and cooperative driving settings to trigger extensive fatal road accidents seem realizable.
\end{itemize}

        \subsubsection{Downward Counterfactual DF Narrative $E'_{c_1}$ }
        
       \begin{itemize}
	\item \textbf{\textit{Designer Goals:}} Although automated peer pressure refers to an agentless self-perpetuating mechanism that emerged through AI-empowered (micro-)targeting\footnote{As for instance successfully performed in the Cambridge Analytica case~\cite{jakubowski2019s}.} on social media, its origins can certainly be traced back to the original benign or neutral economic intentions underlying the early design of social media platforms. Psychologist Richard Freed called present-day social media an\textit{ ``attention economy"}~\cite{lieber} and it is plausible that social media profits from the maximization of utilization time spent by their users.
	\item \textbf{\textit{Knowledge Gaps:}} Early social media designers may not have foreseen the far-reaching consequences of the designed socio-technological artefacts including threats of lethal dimension or even existential caliber according to some present-day viewpoints~\cite{sawers}.
	
	\item \textbf{\textit{Unintended Failures:}} The more attention users pay to social media contents, the more time they may spend with like-minded individuals (consistent with homophily~\cite{jakubowski2019s,prier2017commanding}) and the more they may be prone to automated peer pressure. The latter can an also be partially fueled by social bots aggravating polarization~\cite{yan2020asymmetrical}. The bigger the success of information warfare and targeted disinformation on social media and the higher the performance of the AI technology empowering it, the more groups of like-minded peers could (but of course not necessarily) uptake misleading ideas. Individuals could then -- via these repercussions -- sense \textit{a social pressure to suppress their critical thinking} and get accustomed to simply copy in-group narratives irrespective of their contents. This scenario could in turn play into the hands of malicious actors of the type mentioned in~\ref{ea1} and raise the amount and intensity of the lethal and catastrophic scenarios of the sort described in~\ref{aa3}. 
	
\end{itemize}

           \subsubsection{Downward Counterfactual DF Narrative $F'_{d_1}$ }   
    
        	     \begin{itemize}
        		\item \textbf{\textit{Designer Goals:}}    Implementation of high-performance AI.
        			\item \textbf{\textit{Knowledge Gaps:}} Designers cannot predict the emergence of yet unknown global risks for which no scientific explanatory framework exists (otherwise that would contradict the fundamental unpredictability of future knowledge creation mentioned in Subsection~\ref{arcra}). Given that the past does not contain data patterns of yet never instantiated hazards, the datasets utilized to train "high-performance" AI cannot already have these eventualities reflected in their metrics.
	\item \textbf{\textit{Unintended Failures:}} Exemplary failures that resulted from this unavoidable type of knowledge gap, are multiple post-COVID AI performance issues~\cite{lamb,rahman2020adversarial,vieve}. Simultaneously, humanity relies more and more on medical AI systems. Would humans have been confronted with a more aggressive type of yet unknown biological hazard requiring even faster reactivity, it is conceivable that under the wrong constellations, the AI systems optimizing on metrics pertaining to the then deprecated old or on the novel but yet too scarce and thus biased datasets~\cite{vieve} could have led to unreliable policies up to the potential of a major risk.
\end{itemize}

\section{Discussion}
\label{dis}

\subsection{ Hybrid Cognitive-Affective AI Observatory -- Transdisciplinary Integration and Guidelines}
\label{nearterm}

In this Subsection~\ref{nearterm}, we compile near-term AI safety guidelines with respect to: 1) the factual RDA clusters introduced in Section~\ref{da} and 2) the RDA-based RCRA clusters from Subsection~\ref{rcraex}. For 2), we only specify the necessary supplementary and non-overlapping guidelines to avoid repetitions.

\subsubsection{Near-term Guidelines for Risks \textit{Ia} and \textit{Ib}}
\label{near1}

\paragraph{\textbf{RDA:}}
\begin{itemize}
	\item $ A_{a_1} $: Clearly, for risk \textit{Ia} instances of \textit{adversarial cluster 1} related to the misuse of generative AI to facilitate cybercrimes (e.g.\ via impersonation within social engineering phone calls), already known security measures regarding identity check are needed as minimum requirement. A standard approach to mitigate dangers of malevolent impersonation~\cite{yampolskiy2008mimicry} is to go beyond something you are (biometric)~\cite{yampolskiy2010taxonomy}, and to also require something you know (password)~\cite{yampolskiy2006analyzing} and/or something you have (ID card). Generally, an awareness-raising training of users and employees on social engineering methods including the novel combination possibilities emerging from malicious generative AI design seems indispensable. In addition, it may be helpful to systematically complement those measures with old-fashioned but potentially effective pre-approved but updatable private arrangements made \textit{offline} which can also employ offline elements for identity check. For instance, the malicious actor may not be able to react appropriately in real-time if presented with a from his perspective semantically unintelligible inspection question making use of offline pre-agreed upon (dynamically updated) linguistically ciphered insider idioms. The induced confusion could consequently help to dismantle the AI-aided impersonation attempt.
	 Having said this, it is important to analyze the attack surface that the availability of voice cloning and even video impersonation with generative AI brings about when instrumentalized for attacks against widespread voice-based or video-based authentication methods.
	\item $ A_{a_2} $: This cluster pertaining to AI-aided defamation, harassment, revenge and sextortion exhibits the need for far-reaching legislatures for the protection of potential victims. Legal frameworks but also social media platforms may need to counteract large-scale propagation of material that threatens the safety of targeted entities. Social services could initiate emergency call hotlines for dangerous deepfake victimization. Moreover, the creation of local temporary shelters or havens combining a team of transdisciplinary experts and volunteers for acute phases immediately succeeding the release of compromising material on social media channels appears recommendable. However, the initiation of a societal-level debate and education could foster destigmatization of deepfake instrumentalized for defamation, harassment and revenge. It could dampen the effects of widely distributed compromising material once the general public looses interest in such currently salient elements. More broadly, educating the public about the capabilities of deep-fake technology could be helpful in mitigating defamation, harassment and sextortion since just like society learned to deal with fake Photoshop images, society can also learn scepticism towards AI-generated content. 
		\item $ A_{a_3} $: AI-aided misinformation and disinformation represents a highly complex socio-psycho-technological threat landscape that needs to be addressed at multiple levels using  multi-layered~\cite{whyte2020deepfake} approaches. For instance, in a recent work addressing the malicious applications of generative AI and corresponding defenses, Boneh et al.~\cite{boneh2019relevant} provide a list of directly or indirectly concerned actors: \textit{"authors of fake content;
		authors of applications used to create fake content; owners
		of platforms that host fake content software; educators who
		train engineers in sensitive technologies; manufacturers and
		authors who create platforms and applications for capturing content (e.g., cameras); owners of data repositories used to
		train generators; unwitting persons depicted in fake content
		such as images or deepfakes; platforms that host and/or distribute
		fake content; audiences who encounter fake content;
		journalists who report on fake content; and so on"}. Crucially, as further specified by the authors, \textit{``a precise threat model capturing
	the goal and capabilities of actors relevant to the system
being analyzed is the first step towards principled defenses"}~\cite{boneh2019relevant}. In fact, as briefly adumbrated in Subsection~\ref{rcraex}, the format of the RDA-based RCRA-DFs we proposed for risk \textit{Ia} and \textit{Ib} was purposefully instantiating exactly that -- a threat model. Overall, we thus recommend grounding the development of near-term AI safety defenses (as applied to AI-aided disinformation but also more generally) in RDA-based RCRA-DFs that can be once generated potentially retroactively diversified by novel DF narrative instances tailored to the exemplary actors mentioned by Boneh and collaborators. This could broaden the RCRA results and allow for an enhanced targeted development of countermeasures.
			\item $ A_{a_4} $: For this AI-aided form of non-consensual voyeurism, the measures of an emergency hotline and a specialized haven as mentioned under cluster $ A_{a_2} $ are likewise applicable. Legislators need to be informed on psychological consequences especially for underage victims. While cluster $ A_{a_2} $ implied the overt public dissemination of compromising material by what minor individuals would be less at risk given the potential repercussions, the purely voyeuristic case can often be \textit{covert} and attracts motivational profiles that can target minor individuals~\cite{Hao}. In addition, it might be valuable to proactively inform the general public and also adult population groups susceptible to this issue in order to lift the underlying taboos and to mitigate negative psychological impacts. In the long run, instantiations of this cluster are unlikely to be prevented any more than one can prevent someone fantasizing about someone else. Hence, in the age of fake generative AI artefacts with the virtualization of fake acts of heterogeneous nature normally violating physical integrity in the real-world, it might become fundamentally important to re-assess and/or update societal notions intimately linked to virtual, physical and hybrid body perception in a critical and open dialogue. 
	
				\item $ A_{a_5} $: With regard to AI-aided espionage, companies and public organizations in sensitive domains need to broadly create awareness especially related to the risk of fake accounts with fake but real appearing profile pictures. For instance, since the generator in a generative adversarial network (GAN)~\cite{goodfellow2014generative} is by design imitating features from a given distribution, advanced results of a successful procedure could appear ordinary and more typical -- potentially facilitating a psychologically-relevant intrinsic camouflaging effect. In effect, according to a recent study focused on the human perception of GAN pictures displaying faces of fake individuals that do not exist, \textit{``[...] GAN faces were more likely to be perceived as real than Real faces"}\footnote{Note that on the long-term this could in theory skew the unconscious internal model internet users exposed to more and more synthetic faces have of how human faces look like. Outliers from the real distribution could be met with more surprise at the subpersonal level. However, the latter might already be the case today with the widespread use of enhancing filters on social-media. }~\cite{tucciarelli2020realness}. Beyond that, the authors described an increased social conformity towards faces perceived as real independently of their actual realness. This is concerning also in the light of the extra cluster $ E_{c_1} $ of automated peer pressure that could make AI-aided espionage easier. A generic trivial but often underestimated guideline that may also apply to AI-aided open-source intelligence gathering would be to reduce the sharing of valuable information assets via social media channels and more generally on publically available sources to a minimum. Finally, to confuse person-tracking algorithms and prevent surveillance, camouflage~\cite{young2019calibration} and adversarial patches~\cite{xu2020adversarial} embedded in clothes and accessoires can be utilized.

		\item $ R_{a_1} $: As deep-fake technology proliferates and is used in numerous criminal domains, it is conceivable that an arms-race between malevolent fakers and AI forensic experts~\cite{baggili2019founding,schneider2020ai} will ensue, with no permanent winner. Given that this cluster $ R_{a_1} $ covers a wide variety of research domains in which security researchers and white hats attempt to preemptively emulate malicious AI design activities to foster safety awareness, a consequential recommendation appears to actively support such research at multiple scales of governance. Talent in this adversarial field would need to be attracted by tailored incentives and should not be limited to a standard sampling from average sought-after skill profiles in companies, universities and public organizations of high social reputation. This may also help to avoid an undesirable drift to adversaries for instance at the level of information operations risking reinforcing capacities mentioned in the downward counterfactual DF narratives on cluster $ A'_{a_3} $, $ R'_{a_1} $ and $ E'_{a_1} $ presented in Subsection~\ref{rcraex}. Hence, a monolithic approach in AI governance with a narrow focus on ethics and unintentional ethical failures is insufficient~\cite{aliman2020hybrid}.  Finally, we briefly address guidelines related to a specific $ R_{a_1} $ issue concerning \textit{science} (as asset of invaluable importance for a democratic society~\cite{rosenberg2013reinvigorating}) that did not yet gain attention in AI safety and AI governance but that makes further inspections appear imperative in the near-term. Namely, targeted studies on \textit{AI-aided deception in science} to produce AI-generated text disseminated as \textit{fake research articles} (see the research prototype developed by Yampolskiy~\cite{TransformerJrManuscript-GPTWSC} in another research context) and possibly AI-generated audiovisual or other material meant to display \textit{fake experiments} or also \textit{fake historical samples} (see the recent MIT deepfake demonstration~\cite{moon} developed for educative purposes). However, this technical research direction requires a supplementation by transdisciplinary experts addressing the socio-psycho-technological impacts and particularly the \textit{epistemic} impacts of corresponding future risk instantiations.  We suggest that for a safety-relevant sense-making, AI governance may even need to stimulate debates and exchanges on the very epistemological grounding of science -- \textit{before} e.g.\ future texts written by maliciously designed sophisticated AI bots (also called sophisbots~\cite{boneh2019relevant}) infiltrate the scientific enterprise with submissions that go undetected. For instance, there is a \textit{fundamental discrepancy}\footnote{	Bayesian and empiricist epistemological stances placing the \textit{empirical collection of evidence} and the identification of \textit{true beliefs} at the center of science may link AI-aided deception to \textit{``epistemic threats"}~\cite{fallis2020epistemic} -- knowledge-relevant impairments of belief-updating which they already see emerging via deepfakes (i.a.\ subsuming a general decrease of information in audiovisual samples~\cite{fallis2020epistemic}).  By contrast, Popperian epistemic views~\cite{popper2014conjectures} and especially their Deutschian extension~\cite{deutsch2011beginning} predominantly emphasize in the first place the \textit{explanatory} and \textit{criticism-centered} purpose of science next to the (experimental) \textit{falsifiability} of hypotheses. Strikingly, Deutsch describes science as the endless quest for invariant, \textit{hard-to-vary explanations} of reality~\cite{deutsch2011beginning}. On this view, AI-aided deception in science may be practically problematic, but without question solvable. In fact, while the empiricist direction faces epistemic threats and a post-truth difficulty, the Popperian and Deutschian direction may neither see explanatory knowledge, truth, falsifiability nor the scientific method per se at risk.} between how Bayesian and empiricist epistemology would analyze this risk vs.\ how Popperian critical rationalist epistemology would view the same risk. Disentangling this epistemic issue is of high importance for AI safety and beyond as becomes apparent in the guidelines linked to the next cluster $ E_{a_1} $ below.
	
		\item $ E_{a_1} $: Near-term guidelines to directly tackle this extra cluster associated to automated disconcertion seem daunting to formulate. However, as a first small step, one could focus on how to avoid exacerbating it. One reason why this cluster may seem difficult to address is due to its deep and far-reaching \textit{epistemic} implications pertaining to the nature of falsification, verification, fakery and (hyper-)reality~\cite{baudrillard1994simulacra} itself. With regard to this feature of epistemic relevance, $ E_{a_1} $ exhibits a commonality with the just introduced different risk of AI-aided deception in science. We postulate that in the light of pre-existing fragile circumstances in the scientific enterprise including the emergence of modern "fake science"~\cite{hopf2019fake} patterns but also the mentioned fundamental discrepancies across epistemically-relevant scientific stances, AI-aided deception in science could have direct repercussions on automated disconcertion. First, it could for instance unnecessarily aggravate automated disconcertion phenomena in the general public as e.g.\ the belief in \textit{epistemic threats}~\cite{fallis2020epistemic} could increase people's subjective uncertainty. Second, a reinforced automated disconcertion can subsequently be weaponized and instrumentalized by malicious actors with lethal consequences as generally depicted under the downward counterfactual DF narrative  $ E'_{a_1} $ described in Subsection~\ref{rcraex}. This explains our near-term AI governance recommendation to address AI-aided deception in science as transdisciplinary collaborative endeavor analyzing socio-psycho-technological and epistemic impacts.

			\item $ R_{b_1} $: For this cluster linked to risk \textit{Ib} and pertaining to research on AI vulnerabilities currently performed by security researchers and white hats, we recommend (as analogously already explained in $ R_{a_1} $) to recruit such researchers preemptively. In this vein, Aliman~\cite{aliman2020hybrid} proposes to \textit{``organize a digital security playground where
			"AI white hats" engage in adversarial attacks against AI architectures and share
			their findings in an open-source manner"}. For the specific domain of intelligent systems, it is advisable to proactively equip these AIs with technical self-assessment and self-management capabilities\footnote{The conjunction of technical self-assessment and self-management has been summarized under the synonymous umbrella terms of \textit{Type I AI self-awareness}~\cite{aliman2020hybrid}, \textit{self-awareness functionality}~\cite{aliman2019orthogonality} or simply \textit{self-awareness}.}~\cite{aliman2019orthogonality} allowing for better real-time adaptability for the eventuality of attack scenarios known from past incidents or proof-of-concept use cases studied by security researchers and white hats. However, it is important to keep in mind that challenges from this cluster also deal with zero-day AI exploits, they are the unknown unknowns and cannot be meaningfully anticipated and prevented, though it is realized that many issues could be caused by under-specification in machine learning systems~\cite{d2020underspecification}.
	
\end{itemize}

\paragraph{\textbf{RCRA (additional non-overlapping guidelines):}}
\begin{itemize}
	\item $ A'_{a_2} $: Generally, one possible way to systematically reflect upon defense methods for specific RCRA instances (generated from \textit{downward} counterfactual clusters) of harm intensity $ h_{down}\geq \tau$, could be to perform corresponding \textit{upward} counterfactual deliberations targeting a harm intensity $ h_{up}< \tau$. As briefly introduced in Subsection~\ref{arcra}, upward counterfactuals refer to those ways in which a certain event could have turned out better but did not. Recently, Oughton et al.~\cite{oughton2019stochastic} applied a combination of downward and upward counterfactual stochastic risk analyis to a cyber-physical attack on electricity infrastructure. In short, the difference to the method that we propose is that instead of focusing on \textit{slightly better upward counterfactuals given the factual event} as made sense in the case of Oughton et al., we suggest a threshold-based selection of \textit{{below threshold upward counterfactuals given above threshold downward counterfactuals}\footnote{Since as mentioned earlier, lower harm intensity may lead to more perceiver-dependent differences, one does not exactly need to establish which exact intensity, one only needs to know \textit{that} it is a non-lethal upward counterfactual scenario.}}.  For instance, as applied to the present downward counterfactual cluster $ A'_{a_2} $ which also included a narrative instance describing suicide attempts with lethal outcomes as a consequence of AI-aided defamation, harassment and revenge, it could simply consist in \textit{deliberations on how to avoid these lethal scenarios}. This could be implemented by deliberating from the perspective of planning a human, hybrid or fully automated AI-based emergency team response with a highly restricted timeframe (e.g.\ to counteract the domino-effect initiated by the deployment of the deepfake sample on social media). Next to a proactive combination of deepfake detectors and content detectors for blocking purposes that can fail, a reactive automated social network graph analysis AI combined with sentiment analysis tools could be trained to detect large harassment and defamation patterns that if paired with the sharing of audiovisual samples, can prompt a human operator. This individual could then decide to call in social services that in turn proactively contact the target offering support as analogously mentioned under the guidelines for the factual RDA sample $ A_{a_2} $.

\item $ A'_{a_3} $: For this downward counterfactual cluster on AI-aided misinformation and disinformation of at least lethal dimensions, we focus on recommendations pertaining to journalism-relevant defenses and bots on social media. Disinformation from fake sources could be counteracted with the use of blockchain-based reputation systems~\cite{almasoud2020smart} to assess the quality of information sources. Journalists could also entertain a collective blockchain-based repository containing all news-relevant audiovisual deepfake samples whose authenticity has been refuted so far. This tool could be utilized as publically available high-level filter to evade certain techniques of disinformation campaigns. Moreover, the case of hazardous large-scale disinformation supported by sophisticated automated social bots is of high relevance for what one can term \textit{social media AI safety}. Ideally, \textit{tests} for a "bot shield" enabling some bot-free social media spaces could be crafted. However, it is conceivable that at a certain point, AI-based bot detection~\cite{cresci2020decade} might become futile. Also, social bots already fool people~\cite{cresci2017paradigm,yan2020asymmetrical} and many assume that humans will become unable to discern them in the long-term. Nevertheless, it could be worthwhile viewing what one could have done better already with present technological tools (the upward counterfactuals) -- which can also include the consideration of divergent unconventional solutions or novel formulations of questions. As stated by Barrett,\textit{ ``[...] progress in science is often not answering old questions but asking better ones"}~\cite{barrett2017theory}. Perhaps, in the future, humans could still devise bot shielding tactics that could attempt to bypass epistemic issues~\cite{alimansw2020} intrinsic to imitation game and Turing Test~\cite{turing1950computing} derivatives where "real" and "fake" become relative.

\item $ A'_{a_4} $: To tackle suicidal ideation as a consequence of AI-aided non-consensual voyeurism that enters the awareness of the targeted individual, one may need to extend the countermeasures already mentioned in the factual RDA counterpart $ A_{a_4} $ of this cluster (which also included the creation of public awareness and the removal of associated taboos). Social services and public institutions like universities and schools could offer emergency psychological interventions for the person at risk. Next to necessitated measures at the level of legal frameworks to protect underage victims, the subtle case of adult targets calls for instance for a civil reporting office collaborating with social media platforms which could initiate a critical dialogue with the other party to bring about an immediate deletion or at least categorical refraining from further dissemination of the material which can be calibrated to the expectations of the target. Recently, the malicious design of deepfakes has been described as a \textit{``[...] serious threat to psychological security"}~\cite{pantserev2020malicious}. Adult targets may despite the synthetic nature of the deepfake samples and often eventually their private character restricted to a personal possession of the agent in question, perceive their mere existence as  degradation~\cite{ohman2019introducing} -- a phenomenon certainly requiring social discourses in the long-term. For a principled analytical approach, an extensive psychological research program integrating a collaboration with i.a.\ AI security researchers could be helpful in order to be able to contextualize relevant socio-psycho-technological aspects against the background of advanced technical feasibility. Importantly, instead of limiting this research to deepfake artefacts in the AI field, one needs to also cover novel hybrid combination possibilities available for the design of non-consensual voyeuristic material. Notably, this includes blended applications at the intersection of AI and virtual reality~\cite{aliman2020mal,cole} (or augmented reality~\cite{hybri}).

		\item $ R'_{a_1} $:  Concerning proactive measures against future research where an adversary designs self-owned intelligent systems to trigger lethal accidents on victim intelligent systems, one might require legal norms setting minimum requirements on the techniques employed for the cybernetic control of systems deployed in public space. From an adversarial AI perspective, this could include the obligation to integrate regular updates on AI-related security patches developed in collaboration with AI security researchers and white hats that also study advanced physical adversarial attacks. This becomes particularly important as many stakeholders are currently unprepared in this regard~\cite{kumar2020adversarial}. As guideline, we propose that future adversarial AI research endeavors explore attack scenarios where adversarial examples on physically deployed intelligent systems are delivered by another physically deployed intelligent system which potentially offers more degrees of freedom to the malicious actor.  From a systems engineering perspective, any intelligent system might need to at least integrate multiple types of sensors and check for inconsistencies at the symbolic level. Next to explainability requirements, a further valuable feature to create accountability in the case of accidents could be a type of self-auditing via self-assessment and self-management~\cite{aliman2019orthogonality} allowing for a retrospective counterfactual analysis on what went wrong.
\item $ E'_{a_1} $: As its factual counterpart $ E_{a_1} $, this counterfactual cluster $ E'_{a_1} $ refering to automated disconcertion instrumentalized for AI-aided information warfare and agitation on social media with the risk to incite lethal violence at large scales, represents a weighty challenge of international extent. As for $ E_{a_1} $, multi-level piecemeal tactics of constructive small steps such as e.g.\ targeted methods to avoid exacerbating it may be valuable. Concerning AI governance, that could include the strategies mentioned under $ E_{a_1} $ but also more general efforts in line with international frameworks that aim to foster strong institutions and error-correction via life-long learning (see e.g.~\cite{Delphi2} for an in-depth discussion).

			\item $ R'_{b_1} $: For this counterfactual cluster pertaining to malicious research on vulnerabilities of deployed AI systems with the goal to trigger extensive fatal road accidents, we recommend tailored measures analogous to those presented for the counterfactual cluster $ R'_{a_1} $. 

\end{itemize}
\subsubsection{Near-term Guidelines for Risks \textit{Ic} and \textit{Id}}
\label{near2}
As can already be realized from the scope of the AI safety guidelines proposed in Subsection~\ref{near1} which are grounded in our AI observatory exemplification of RDA and RCRA, modern AI technology cannot be analyzed in isolation. In our view, due to the complex multi-causal socio-psycho-technological interwovenness underlying AI risks and their instantiations, AI safety requires an inherently \textit{transdisciplinary}, \textit{hybrid} and \textit{cognitive-affective} approach~\cite{aliman2020hybrid}. Transdisciplinarity is especially required to avoid cognitive blind spots within AI safety risk analyses and formulations of countermeasures or guidelines. AI safety needs a hybrid perspective to incorporate the intricacies of human-computer interactions necessitating a consideration of \textit{human nature} next to purely technological viewpoints. Finally, a cognitive-affective perspective is called for due to the inseparably affective nature of human cognition~\cite{barrett2015interoceptive,kleckner2017evidence} whose disregard in AI development can consequently engender significant safety issues by virtue of a lack of requisite variety~\cite{aliman2019requisite}. While the last Subsection~\ref{near1} focused on guidelines concerning the AI risks \textit{Ia} and \textit{Ib} related to intentional malice, this Subsection~\ref{near2} is linked to the risks \textit{Ic} and \textit{Id} related to mistakes and unintentional failures which are often of ethically-relevant nature. This specific avenue of research represents a well-studied field at the core of modern AI ethics which recognizes multidisciplinarity, human-centeredness and socio-technical contextualization as important requirements~\cite{dignum2020ai}. In the last years, a large multiplicity of heterogeneous AI ethics guidelines have been proposed at an international level~\cite{floridi2019establishing,hagendorff2020ethics,mittelstadt2019ai,whittlestone2019role}. We refer the reader to Jobin et al.~\cite{jobin2019global} for a global overview of internationally proposed AI ethics guidelines which are directly of relevance for the 5 failure clusters ($ F_{c_1} $ to $ F_{c_5} $) linked to risk \textit{Ic} from the RDA presented in Subsection~\ref{rda2}. In the following, we focus on the few remaining RDA and RCRA clusters which are not classically in the primary focus of AI ethics.
\paragraph{\textbf{RDA:}}
\begin{itemize}

	\item $ E_{c_1} $: This cluster related to automated peer pressure can be i.a.\ met by measures raising public awareness on the dangers of the confirmation bias~\cite{gu2014research,yoo2007ideological} reinforced via AI-empowered social media. However, a possible upward counterfactual on that issue would be to revert negative consequences of automated peer pressure by utilizing it for beneficial purposes. For instance, it is cogitable that automated peer pressure need not represent a threat would it simply perhaps paradoxically socially \textit{reinforce critical thinking} instead of reinforcing tendencies to blindly copy in-group narratives. Ideally, such a peer pressure would reinforce  \textit{heterophily} (the antonym of homophily) with regard to various preferences with one notable exception being the critical thinking mode itself. Hence, one interesting future-oriented solution for AI governance may be education and life-long learning~\cite{Delphi2} conveying critical thinking and criticism as invaluable tools for youth and general public. For instance, critical thinking skills fostered in the Finnish public education
	system were effective against disinformation operations~\cite{jakubowski2019s}. In fact, critical thinking, criticism and transformative contrariness may not only represent a strong shield to tackle disinformation or automated disconcertion and its risk potentials (cluster $ E_{a_1} $ and $ E'_{a_1} $ respectively), but it also represents a crucial momentum for human creativity~\cite{aliman2020hybrid,tsao2019creative}. Generally, peer pressure is in itself a psychological tool that could be systematically used for good, for example by creating an artificial crowd~\cite{yampolskiy2012wisdom} of peers with all members interested in desirable behaviors such as education, start-ups or effective altruism. A benevolent crowd of peers can then counteract hazardous bubbles on social media. 

	\item $ F_{d_1} $: Concerning AI failures rooted in \textit{unanticipated} and yet unknown post-deployment scenarios, it becomes clear that accuracy and other AI performance measures cannot be understood as conclusive and engraved in stone. A possible proactive measure against post-deployment instantiations of yet unknown AI risks could be the establishment of a generic corrective mechanism. Problems which AI systems experience during its deployment due to differences between training and usage environments can be reduced via increased testing and continued updating and learning stages. On the whole, multiphase deployment, similar to vaccine approval phases, can reduce an overall negative impact on society and increase reliability. Finally, for each safety-critical domain in which AI predictions are involved in the decision-making procedure, one could -- irrespective of present-day AI performance -- foresee the proactive planification of a human response team in case of sudden expanding anomalies that a sensitized and safety-aware human operator could detect.

\end{itemize}

\paragraph{\textbf{RCRA (additional non-overlapping guidelines):}}
\begin{itemize}

	\item $ E'_{c_1} $: A twofold guideline for this counterfactual cluster (refering to automated peer pressure with lethal consequences via automated disconcertion instrumentalized for AI-aided disinformation), could be to weaken the influence of social bots by measures described under cluster $ A'_{a_3} $ and by transforming automated peer pressure into strong incentives for critical thinking as stated in $ E_{c_1} $.
	
	\item $ F'_{d_1} $: Finally, for this cluster of major risk dimension being the counterfactual counterpart of cluster $ F_{d_1} $, we emphasize the importance of an early proactive response team formation in contexts such as for instance medical AI, AI in the financial market, AI-aided cybersecurity and critical intelligent cyberphysical assets. In short, AI systems should by no means be understood to be able to truly operate independently in a given task even if current excellent performance measures seem to suggest so. In the face of unknown and unknowable changes, performance is a moving target which if mistaken as conclusive and static could endanger human lifes.
\end{itemize}

\subsection{Long-Term Directions and Future-Oriented Contradistinctions}
\label{longterm}

After having introduced a broad variety of near-term guidelines for future AI observatory endeavors based on the exemplified systematic factual and counterfactual retrospective analyses, we provide a differentiated more \textit{general} outlook on explicitly \textit{long-term} AI safety directions. For this purpose, we select two recent theoretical AI safety paradigms: on the one hand a direction that has been termed \textit{artificial stupidity} (AS) (see~\cite{guideyam,trazzi2018building,trazzi2020artificial}) and on the other hand, a direction that we succinctly call \textit{eternal creativity} (EC) stemming from recent work~\cite{aliman2020artificial,aliman2020error,aliman2020hybrid}. Thereby, note that these two paradigms are by no means postulated to represent  the full panoply of nuances and views across the entirety of the young AI safety field. Rather, we select these specific two examples because critical contradistinctions ascertainable via a comparative analysis point to a set of decisive bifurcations which might be of particular interest for the AI safety community due to their potentially \textit{axiomatic} relevance for the future of AI research. While AS and EC coincide in multiple short-term considerations given their common \textit{hybrid cognitive-affective} nature and their emphasis on \textit{cybersecurity-oriented} practices, they fundamentally differ with regard to 3 future-relevant contradistinctions.    \par
We consider the following 3 contradistinctive leitmotifs: 1) \textit{regulatory distinction criterium}, 2) \textit{regulatory enactment} and 3) \textit{substrate management}. First, while AS primarily considers \textit{intelligence} levels for 1), EC ponders the ability to consciously create and understand \textit{explanatory knowledge}. Second, whilst AS foresees deliberate \textit{restrictions} of AI capabilities as tool for 2), EC especially tackles their systematic \textit{enhancement}. Third, while AS views \textit{substrate-dependent} hardware analyses (next to software considerations) for bounded equalization between humans and AIs as approach to 3), EC aims at unbounded \textit{substrate-independent} functional augmentation. While there exist certainly more possible lines along which one could compare AS and EC, we focus on the mentioned 3 themes due to their urgency and potential to foster constructive dialectics in future theoretical and applied AI (safety) research beyond AI observatory contexts. In Subsection~\ref{as} and~\ref{ec}, we briefly provide a general introduction followed by a summarization of long-term AI safety guidelines formulated from the perspective of AS and EC respectively as seen through the lens of these 3 contradistinctions.
\subsubsection{Paradigm Artificial Stupidity (AS)  }
\label{as}

One core assumption in the AS paradigm is that an artificial general intelligence (AGI) \textit{"[...] can be made safer by limiting its computing
power and memory, or by introducing Artificial
Stupidity on certain tasks"}~\cite{trazzi2018building}. Thereby, an AI system is understood to be made \textit{artificially stupid} on a certain task if its capabilities are deliberately limited by human designers for the purpose of matching the human performance on that task. One mentioned exemplary domain where such a technique is already applied is in text-to-speech synthesis such as e.g\ in Google Duplex, an AI for natural conversations over the phone whose implementation included \textit{"[...] the incorporation of speech disfluencies (e.g. "hmm"s and "uh"s)"}~\cite{49194}. Another example is the context of video games where AI can in principle vastly exceed human performance which is however purposefully restricted in order to allow for a positive human-centered gaming experience. More generally, there are many AI application domains where it is human-desirable to mimic anthropic performance or behavioral patterns for an improved customer service. These cases correspond to a type of imitation game which only succeeds if the AI does not reveal latent super-human capabilities. From that point of view, the AS paradigm conceives of making an AI artificially stupid as being necessary to making it pass a Turing test~\cite{trazzi2020artificial,trazzi2018building}.  \par Simultaneously, in the last years, AI achieved superhuman-level performance across more and more tasks. Further, it is assumed in AS that \textit{"[...] AI tends to
quickly achieve super-human level of performance
after having achieved human-level
performance}"~\cite{trazzi2020artificial}. Against this background, AS argues distinguishingly that \textit{"[...] by limiting an AI's ability to achieve
a task, to better match humans' ability, an
AI can be made safer, in the sense that its
capabilities will not exceed humans' capabilities
by several orders of magnitude"}~\cite{trazzi2020artificial}. In short, AS postulates that \textit{AI ability needs to be upper-bounded by human performance} since it risks to otherwise become \textit{uncontrollable}\footnote{For an in-depth discussion related to AI uncontrollability and unpredictability, see especially~\cite{yampolskiy2020controllability} and~\cite{yampolskiy2019unpredictability} respectively. } once it turns into what Bostrom termed a \textit{superintelligence} -- an intellect exceeding human cognitive performance across \textit{"[...] virtually all domains of interest"}~\cite{bostrom2014superintelligence}. Such a hypothetical future artificial superintelligence is believed to not necessarily be value-aligned with humans (while potentially becoming unintelligible to humans due to the gaps in performance), to be capable of insidious betrayal (a scenario termed \textit{treacherous turn}~\cite{bostrom2014superintelligence}) and to potentially represent a major risk~\cite{baum2017modeling} to humanity.

\begin{itemize}
	\item \textbf{\textit{Regulatory distinction criterium:}}
	In this light, one can extract \textit{intelligence} (or more broadly "performance" or "cogntive performance" across tasks) as the recurring theme of relevance for regulatory AI safety considerations under the AS paradigm. At a first level, one could identify two main safety-relevant clusters: a cluster comprising all AIs that are less or equally capable than an average human~\cite{trazzi2020artificial} and another cluster of superintelligent AI systems. The latter can be further subdivided into three classes of systems as introduced by Bostrom~\cite{bostrom2014superintelligence}: 1) speed superintelligence, 2) collective superintelligence and 3) quality superintelligence. According to Bostrom, the first ones \textit{``can do
	all that a human intellect can do, but
	much faster"}, the second ones are \textit{``composed of a large
number of smaller intellects such that
the system’s overall performance across
many very general domains vastly outstrips
that of any current cognitive system"} and the third ones are \textit{``at least as fast as a human
mind and vastly qualitatively smarter"}~\cite{bostrom2014superintelligence}. 
	
	\item \textbf{\textit{Regulatory enactment:} } In a nutshell, AS recommends \textit{limiting} an AI in hardware and software such that it does not attain any  of these enumerated sorts of superintelligence since \textit{"[...] humans could lose control over the AI"}~\cite{trazzi2018building}. AS foresees regulatory strategies on \textit{``how to
		constrain an AGI to be less capable than
		an average person, or equally capable,
		while still exhibiting general intelligence"}~\cite{trazzi2020artificial}.

	\item \textbf{\textit{Substrate management:}}  To limit AI abilities while maintaining functionality, AS proposes multiple practical measures at the hardware and software level. Concerning the former it proposes diverse restrictions especially pertaining to memory, processing, clock speed and computing~\cite{trazzi2020artificial}. With regard to software, it foresees necessary limits on self-improvement as well as measures to avoid treacherous turn scenarios~\cite{trazzi2018building}. Another guideline consists in deliberately incorporating known human cognitive biases in the AI system. More precisely, AS postulates that human biases \textit{{"can limit
			the AGI's intelligence and make the AGI fundamentally
			safer by avoiding behaviors that might harm humans"}}~\cite{trazzi2018building}. Overall, the substrate management in AS can be categorized as \textit{substrate-dependent} because the artificial substrate is among others specifically tuned to match hardware properties of the human substrate for at most equalization purposes. In summary, AS suscribes to the viewpoint that AI safety aims to \textit{"limit aspects of memory, processing, and speed in ways that align with human capabilities and/or prioritize human welfare, cooperative behavior, and service to humans"}~\cite{guideyam} given that AGI \textit{"[...] presents a risk to humanity"}~\cite{guideyam}.

\end{itemize}

\subsubsection{Paradigm Eternal Creativity (EC)}
\label{ec}
According to Deutsch, \textit{``the only uniquely significant thing about humans [...] is our ability to create new explanations [...]"}~\cite{deutsch2011beginning}. He further specifies that explanatory knowledge \textit{"gives people a power to transform nature which is ultimately not limited by parochial factors, as all other adaptations are, but only by universal laws"}~\cite{deutsch2011beginning}. Instead of emphasizing levels of intelligence or of performance across a wide set of tasks when analyzing AI safety issues, EC focuses epistemologically on one unique "task": \textit{the ability to consciously create and understand explanatory knowledge}. Thereby, in EC, explanatory knowledge creation also implies the capability to \textit{consciously} understand. Given that core affect is understood as a fundamental property of consciousness~\cite{barrett2015interoceptive,barrett2017theory} and is linked to cognitive-affective counterfactual deliberations~\cite{aliman2020error}, this excludes philosophical zombie themes~\cite{friston2018self}. (In modern embodied and enactive cognition frameworks~\cite{barrett2017theory,bruineberg2018anticipating}, consciousness is linked to processes of inference for the cybernetic control of a substrate in an environment connected to allostasis~\cite{kleckner2017evidence} (anticipation of needs before they occur~\cite{barrett2017theory}) -- integrating predictions and error signals from external and internal milieu. It is on such cybernetic control grounds that affective dynamics give rise to the egocentric virtual first-person perspective of the world~\cite{rudrauf2017mathematical,williford2018projective} familiar to humans and lacking in present-day AI.)\par

Note that EC's focus on consciously creating and understanding explanatory knowledge is by no means an anthropomorphic assumption forced on AI systems. As elucidated in constructor theory~\cite{deutsch2013constructor,deutsch2015constructor}, a novel explanatory framework in physics, explanatory knowledge creators (of which currently only humans are known) are brought to the fore in physics in an entirely non-anthropocentric way. To put it very simply, constructor theory focuses on \textit{possible} vs.\ \textit{impossible} counterfactuals i.e.\ 
what \textit{could} happen given physical laws and \textit{why} (instead of predictions based on laws of motion and initial conditions). On contemplating the set of all physical transformations that would be \textit{possible} in the universe i.e.\ those that \textit{could} happen, one would notice that the size of the very subset containing those transformations that \textit{actually happen} can be strongly influenced by entities able to create and understand knowledge on how to bring them about~\cite{deutsch2011beginning}. This is how explanatory knowledge creation enters \textit{"the cosmic scheme of things"}~\cite{deutsch2011beginning} and this is also why EC prioritizes the conscious understanding and creation of explanatory knowledge via creativity\footnote{From a psychological and neurocognitive perspective, EC currently views creativity as a tri-partite evolutionary affective construct with varying degrees of \textit{sightedness}~\cite{aliman2020artificial} instead of a \textit{blind} evolutionary process without a goal akin to biological evolution -- as mistakenly assumed by Popper~\cite{aliman2020error,dietrich2015creativity}. This is epistemologically relevant because ideas are \textit{not} created by blind trial and error (as variation and selection in biological evolution). Even if novel idea contents are fundamentally unpredictable a priori, idea variation is partially guided by previous experience, the task and contextual cues i.e.\ there is a non-zero coupling between variation and selection~\cite{dietrich2015creativity}. Creativity itself could have historical roots in \textit{serendipity} and multi-purpose socially \textit{shared doubt}~\cite{aliman2020hybrid} facilitating in theory error-correction but initially largely used to maintain traditions.} instead of intelligence.\par At first sight, given the fundamental unpredictability of future explanatory knowledge, it might seem dangerous for AI safety. Deutsch mentions that \textit{"no good explanation can predict the outcome, or the probability of an outcome, of a phenomenon whose course is going to be significantly
affected by the creation of new knowledge"}~\cite{deutsch2011beginning} and further that this fundamental limitation is something that \textit{"when planning for
the future, it is vital to come to terms with it"}~\cite{deutsch2011beginning}. EC agrees.   EC recently formulated the \textit{AI safety paradox}~\cite{aliman2020error,aliman2020hybrid} stating that value alignment and control are conjugate requirements in AI safety. This means that both prevailing ideals cannot be simultaneously fulfilled. EC also states that \textit{"the price of security is eternal creativity"}~\cite{aliman2020hybrid}. So despite the AI safety paradox, a cybersecurity-oriented and risk-centered AI safety is possible -- when reframed \textit{"as a discipline which proactively addresses AI risks and reactively responds to occurring instantiations of AI risks"}~\cite{aliman2020hybrid}. In short, AI safety is not condemned, it just needs to come to terms with the compulsion to keep correcting and creating solutions "ad infinitum".

\begin{itemize}
\item \textbf{\textit{Regulatory distinction criterium:}} EC distinguishes two  \textit{substrate-independent} and disjunct sets of systems: Type I and Type II systems. Type II systems are all systems for which it is possible to consciously create and understand explanatory knowledge. Type I systems are all systems for which this is an impossible task~\footnote{EC could be stated to apply a constructor-theoretic distinction to AI safety insofar as it applies a possibility-impossibility dichotomy~\cite{deutsch2013constructor} embedded in an explanatory framework to it.}. Thereby, a subset of Type I systems can be conscious (such as non-human mammals) and requires protection akin to animal rights. Obviously, \textit{all present-day AI systems are of Type I} and  \textit{non-conscious}. Type II AI is non-existent today.

	\item \textbf{\textit{Regulatory enactment:} }  
	 In theory, with a Type II AI, \textit{"a mutual value alignment might be achievable via a co- construction of novel values, however, at the cost of its predictability"}~\cite{aliman2020hybrid}. As with all Type II systems (including humans), the future contents of the knowledge they will create are fundamentally unpredictable -- irrespective of any intelligence class\footnote{Under EC, superintelligence is as explained not of distinctive interest. It is also viewed as \textit{not} implying profound \textit{qualitative} differences to human baselines. Following Deutsch, it would be \textit{"[...] subject only to limitations of speed or memory capacity, both of which can be equalized by technology"}~\cite{brockman2020possible}. EC views human augmentation as valid transformative defense strategy~\cite{aliman2020artificial}.}. In EC, this signifies that: 1) \textit{Type~II AI is uncontrollable and requires rights on a par with humans}, 2) \textit{Type II AI could engage in a mutual bi-directional value alignment with humans -- if it decides so} and 3) \textit{it would be unethical to enslave Type~II AI}. (Finally, banning Type II AI is a potential loss of requisite variety and does not hinder malicious actors to do so.) By contrast, regarding Type I AI, EC implies that: 4) \textit{Type I AI is controllable}, 5) \textit{Type I AI cannot be fully value-aligned} across all domains of interest for humans due to an insufficient understanding of human morality, 6) \textit{conscious Type I AI is possible and would require animal-like rights} but it is clearly \textit{non-existent} nowadays. 

		\item \textbf{\textit{Substrate management:}} To avoid functional biases~\cite{reynolds2017teams} due to a lack of diversity in information processing, EC opts for a \textit{substrate-independent} functional view. Irrespective of its specific substrate composition, an overall panoply of systems is viewed as \textit{one} unit with diverse functions. Given Type-II-system-defined cognitive-affective goal settings, a systematic function integration can yield complementary synergies. Notably, EC recommends research on substrate-independent functional \textit{artificial creativity augmentation}~\cite{aliman2020artificial} (artificially augmenting \textit{human}
		creativity \textit{and} augmenting \textit{artificial} creativity). For instance, active inference could technically \textit{increase} Type I AI exploratory abilities~\cite{friston2017active,sajid2019active}. Besides that, in Subsection~\ref{matrcraprep}, we apply a functional viewpoint to \textit{augment} RCRA DF generation by human Type II systems for AI observatory purposes.

\end{itemize}

\section{Materials and Methods}
\label{mat}

\subsection{RDA Data Collection}
\label{matrda}

For the collection of RDA samples utilized for illustration purposes in this paper, we undertook a simple keyword-based web search limited to articles in the period between 2018 and 2020. The main queries (with associated boolean operators) that we considered were: "artificial intelligence","AI",  "autonomous", "neural network", "deepfakes", "AI" AND "bias", "AI" AND "failure",  "AI" AND "security", "AI" AND "safety", "AI" AND "attack". While many terms are tailored to the type of keys represented in the taxonomy (\textit{Ia}, \textit{Ib}, \textit{Ic}, \textit{Id}) that served as basis for categorization in the RDA as introduced in Section~\ref{tax}, we also considered utmost \textit{general} queries such as "artificial intelligence" in order to do justice to the eventuality that we might identify a novel entirely unexpected categorization pattern. With other words, we also foresaw the possibility of not yet encountered anomalies while analyzing the results. As briefly mentioned in Subsection~\ref{proc}, such a case would have been assigned to a \textit{generic placeholder key for novel unknown patterns}. It would have called for  further scrutiny and eventually for a future enlargement of the taxonomy. However, as mentioned in Subsection~\ref{proc}, we did not yet identify any novelty of this kind in the discussed RDA. Though, at a lower level, we discovered \textit{atypical} instances of the pre-existing key-determined clusters. We tagged this atypicality by refering to corresponding clusters with the attribute "extra" -- which was the case for the extra cluster of automated disconcertion linked to risk \textit{Ia} and the extra cluster of automated peer pressure connected to risk \textit{Ic}. \par Self-evidently, the underlying search can be performed in a more sophisticated way in future AI observatory projects. First, a broader range of keys and combinations can be strategically devised in the light of RDA and RCRA results from a previous AI observatory iteration. Second, the efforts can be supported by web crawlers~\cite{hernandez2020semantic}. Third, this could be combined with sentiment analysis tools~\cite{singh2020sentiment} to detect negatively polarized texts of interest for an RDA. Fourth, the creation of novel datasets for text classification~\cite{kowsari2019text} could be undertaken for the pre-existing keys of the taxonomy which might however remain insufficient with regard to placeholders for novel patterns. In this vein, we stress the importance of human analysts for a deep semantic understanding requiring explanatory knowledge especially when it comes to the discovery of subtle novel tendencies within superficially similar text sources. Morever, an intense examination of textual material can lead to a further disentanglement of pre-existing clusters -- which could even reveal the need for a broader change of the taxonomic keys. In short, a safety-aware responsible RDA data collection pipeline is not entirely automatable and requires human-level understanding by analysts.

\subsection{ Interlinking RDA-based RCRA Pre-processing and RCRA DFs}
\label{matrcraprep}
As elucidated in Subsection~\ref{proc}, the preparatory procedure generating candidate RCRA clusters based on RDA instances consisted of 4 consecutive steps: 1) \textit{taxonomization}, 2) \textit{analytical clustering}, 3) \textit{brute-force deliberation and threshold-based pruning} and finally 4) \textit{assembly}. Subsequently, these RCRA clusters served as basis to generate RCRA DFs that we exemplified with short RCRA narratives instantiating these clusters as presented in Subsection~\ref{rcraex}. However, for the sake of simplicity, the exact methodological approach to \textit{interlink} the preparatory procedure and the RCRA co-creation DF was not previously characterized. In a nutshell, we utilized a method we call \textit{complementary cognitive co-creation} (CCC). While other methods are thinkable, we encourage considering CCC where possible for reasons described in the next paragraphs. Beforehand, we must specify that purposefully, the set of researchers involved in the preparatory procedure of the RCRA and the set of researchers performing the ensuing RCRA DFs were \textit{disjunct}. For clarity, we refer to the former as \textit{preparatory group} and to the latter as \textit{executive group}. We explain how a complementary collaborative effort between these groups in the form of CCC can increase the\textit{ variety and illustrative power} of RCRA DFs.\par
After applying taxonomization and analytical clustering to the RDA instances, the preparatory group has been described in Subsection~\ref{proc} to perform brute-force deliberation and threshold-based pruning. While a brute-force search could appear suboptimal at first sight, we specifically considered this option in order to allow for the preparatory group to potentially be able to retrospectively \textit{diversify} the generation of instances performed by the executive group  given the RCRA clusters. This becomes possible, because whilst the preparatory group goes through every single available RDA instance, it attempts to generate an above threshold downward counterfactual that if identified can later turn out to be \textit{utile to store}. In short, when a downward counterfactual is successfully generated for a given RDA sample, the preparatory group can not only maintain the RDA sample, but also store the generated downward counterfactual instance for later RCRA augmentation purposes. Thereby, as briefly specified, generic RCRA \textit{clusters} were used instead of specific instances as inputs for the RCRA DFs to avoid overfitting to the idosyncrasies of unique events and possibly generate a broader variety of DF scenarios. In fact, by solely providing RCRA \textit{clusters} to the executive group at the start of the DFs, we avoid a potentially biased negative influence by the narrow instances of the preparatory group that fulfilled a different primary function (namely the identification of above threshold patterns). To recapitulate, the preparatory procedure can be more precisely re-explained as follows: the preparatory group undergoes all 4 consecutive steps with the crucial additional detail that the brute-force deliberation and threshold-based pruning operation \textit{also} includes \textit{the storing of a successfully generated downward counterfactual instance} for each  maintained factual RDA instance. After this pre-DF processing, the preparatory group delivers the RCRA clusters to the executive group which then engages in generating a variety of narratives instances for each obtained cluster. Post-DF, the executive group compares the generated instances with those imagined by the preparatory group pre-DF. All cases that were not yet considered by the executive group\footnote{Note that if given an RCRA cluster, the executive group would not succeed in imagining a corresponding instance for a narrative, there is always at least one back-up instantiation -- which corresponds to the narrative envisaged by the participatory group pre-DF (whose identification represented the precondition for this cluster to exist in the first place).} but were generated by the preparatory group, are concatenated to the now augmented DFs. Duplicates are ignored.\par

This overall sequence of steps presents a theoretical collaborative basis for \textit{an augmentation of co-creation DFs} to which we refer to with CCC. A further tool that may improve the efficacy of CCC is to add a functional viewpoint (i.e.\ related to information processing in a certain context). On closer inspection, it becomes clear why CCC can profit from a functional or/and (neuro-)cognitive~\cite{aggarwal2019impact,den2019neurodiversity,blume1998neurodiversity}  diversity of the partaking researchers. Given that in the human cognitive domain, variety is the norm~\cite{chapman2020neurodiversity} and heterogeneity can provide requisite variety in complex multi-causal dynamic problem domains~\cite{reynolds2017teams} necessitating collective learning~\cite{aggarwal2019impact} and innovation~\cite{chen2019cognitive}, it makes sense to explore this potential. For instance, while the preparatory group can especially profit from individuals that excel at convergent thinking, the executive group may benefit from divergent thinkers. Pre-DF, the preparatory group needs to map from \textit{one} factual \textit{instance} to \textit{one} counterfactual \textit{instance}. In the DF, the executive group maps from \textit{one} counterfactual \textit{cluster} to \textit{many} counterfactual \textit{instances}. The former requires a horizontal integration at a low level of abstraction while the latter requires a vertical integration from a higher to a lower level of abstraction revealing the potential for \textit{complementary} synergies\footnote{For instance, despite possible significant context-dependent~\cite{chapman2020neurodiversity} hindrances, dyadic mismatches~\cite{bolis2017beyond} and disabilities, autistic traits are also paired with enhanced convergent thinking~\cite{abu2020autistic}, detail-rich thinking~\cite{paola2020autism} and higher verbal creativity~\cite{10.3389/fpsyg.2020.559238} while attention deficit hyperactivity disorder traits have been linked to enhanced divergent thinking~\cite{hoogman2020creativity,white2020thinking} and enhanced originality and flexibility~\cite{white2016scope}. Systematically combining these two complementary cognitive profiles under a CCC-oriented approach to RDA-based RCRA-DFs for AI observatory feedback-loops could engender benefits. }. A CCC-based approach combining a preparatory group comprising i.a.\ individuals with a cognitive profile exhibiting strengths in the former and an executive group i.a.\ sampled from a pool of individuals with strengths in the latter could increase efficiency, variety and illustrative power of RDA-based RCRA co-creation DFs -- critical to raise safety-awareness in experts but also in the public.

\section{Conclusions}
\label{conc}
Starting with a \textit{cybersecurity-oriented} fit-for-purpose taxonomy of ethical distinction, we introduced and exemplified a \textit{retrospective descriptive analysis} (RDA) for future AI observatory projects. Subsequently, we elucidated how to craft a complementary \textit{retrospective counterfactual risk analysis} (RCRA) based on downward counterfactuals from the previously extracted factual RDA samples. Motivated by recent work on risk management of hazardous events~\cite{woo2019downward} and the \textit{functional theory of counterfactual thinking}~\cite{roese2017functional} from social psychology, we elaborated on why an RDA-based RCRA may be suitable for risk analyses in a complex multi-causal domain such as AI safety. Thereafter, in the light of the ethical sensitivity of AI risk instantiations, we discussed the use of harm intensity ratings for samples of an AI observatory given the perceiver-dependent, harm-based and dyadic nature of human cognitive templates in morality~\cite{schein2018theory}. For illustrative purposes, we suggested a threshold-based approach focusing the RDA-based RCRA on downward counterfactuals of at least \textit{lethal} dimensions. On the one hand, such a high threshold may engender fewer discrepancies in the moral perception being related to harm. On the other hand, it may simultaneously represent a suitable threshold reinforcing \textit{mortality salience} (i.e.\ the awareness of one's mortality). From the perspective of a relevant socio-psychological theory denoted terror management theory~\cite{greenberg2011terror,solomon2015worm}, mortality salience -- whose elicitation is also conceivable in co-creation design fictions from HCI including virtual reality settings~\cite{chittaro2017mortality} -- may be able to foster safety-awareness and cautionary attitudes~\cite{chittaro2017mortality,shehryar2005terror}. Against the backdrop of the RDA samples collected and our targeted RDA-based RCRA, we formulated the need for inherently \textit{transdisciplinary} and\textit{ hybrid cognitive-affective} AI observatory and AI safety strategies. As guidelines for future work, we compiled a rich variety of tailored multi-level \textit{near-term} solutions.\par Finally, we provided a differentiated general outlook on \textit{long-term }AI safety directions by axiomatically contrasting two disparate recent AI safety paradigms along relevant contradistinctive leitmotifs. More precisely, we contrasted the \textit{artifical stupidity} (AS) paradigm with the \textit{eternal creativity} (EC) paradigm. While AS and EC share a common cybersecurity-oriented and hybrid cognitive-affective stance with regard to multiple near-term AI safety solutions, they differ fundamentally in many future avenues of research. AS offers \textit{intelligence-focused}, \textit{restriction-based} and tailored \textit{substrate-dependent} long-term guidelines. By contrast, long-term EC guidelines bring into focus \textit{conscious explanatory knowledge creation and understanding} and recommend unbounded \textit{functional augmentation} of \textit{substrate-independent} nature.  While AS suggests utilizing human cognitive performance as upper bound for AI capabilities to limit hardware and software parameters, EC takes a cybernetic perspective according to which humans need to jointly augment both human and AI functions -- e.g.\ via a doubly ambiguous artificial creativity augmentation research.  \par

 In a nutshell, we collated retrospective analyses complemented by future-oriented contradistinctions in order to: 1) apprise future AI observatory projects using \textit{concrete examples} from practice and technically plausible above threshold downward counterfactuals, 2) thematizing possibly decisive bifurcations in future AI (safety) research and 3) pointing out the requirement of a constructive collaborative dialectical approach addressing those. As stated by Popper, \textit{"while differing widely in the various little bits we know, in our infinite ignorance we are all equal"}~\cite{popper2014conjectures}. Time might tell whether the assumption that \textit{"the price of security is artificial stupidity"} or rather that \textit{ "the price of security is eternal creativity"}~\cite{aliman2020hybrid} (or none of those) turns out to practically solve long-term AI safety problems. Either way, explanatory knowledge \textit{co-creation} can heavily influence whether we will succeed in \textit{understanding} how to transform today's vulnerability awareness and mortality salience into the currently known or unknowable \textit{upward} counterfactuals of our counterfactual future.
\reftitle{References}


\externalbibliography{yes}
\bibliography{IEEEexample}

\begin{thebibliography}{-------}
\providecommand{\natexlab}[1]{#1}

\bibitem[Amodei \em{et~al.}(2016)Amodei, Olah, Steinhardt, Christiano,
  Schulman, and Man{\'e}]{amodei2016concrete}
Amodei, D.; Olah, C.; Steinhardt, J.; Christiano, P.; Schulman, J.; Man{\'e},
  D.
\newblock {Concrete problems in AI safety}.
\newblock {\em arXiv preprint arXiv:1606.06565} {\bf 2016}.

\bibitem[Dafoe(2018)]{dafoe2018ai}
Dafoe, A.
\newblock {AI governance: a research agenda}.
\newblock {\em Governance of AI Program, Future of Humanity Institute,
  University of Oxford: Oxford, UK} {\bf 2018}.

\bibitem[Everitt \em{et~al.}(2018)Everitt, Lea, and Hutter]{everitt2018agi}
Everitt, T.; Lea, G.; Hutter, M.
\newblock {AGI safety literature review}.
\newblock  Proceedings of the 27th International Joint Conference on Artificial
  Intelligence,  2018, pp. 5441--5449.

\bibitem[Fjeld \em{et~al.}(2020)Fjeld, Achten, Hilligoss, Nagy, and
  Srikumar]{fjeld2020principled}
Fjeld, J.; Achten, N.; Hilligoss, H.; Nagy, A.; Srikumar, M.
\newblock {Principled artificial intelligence: Mapping consensus in ethical and
  rights-based approaches to principles for AI}.
\newblock {\em Berkman Klein Center Research Publication} {\bf 2020}, {\em 1}.

\bibitem[Irving \em{et~al.}(2018)Irving, Christiano, and Amodei]{irving2018ai}
Irving, G.; Christiano, P.; Amodei, D.
\newblock {AI safety via debate}.
\newblock {\em arXiv preprint arXiv:1805.00899} {\bf 2018}.

\bibitem[Turchin \em{et~al.}(2019)Turchin, Denkenberger, and
  Green]{turchin2019global}
Turchin, A.; Denkenberger, D.; Green, B.P.
\newblock {Global Solutions vs. Local Solutions for the AI Safety Problem}.
\newblock {\em Big Data and Cognitive Computing} {\bf 2019}, {\em 3},~16.

\bibitem[{The Agency for Digital Italy}(2020)]{italianAI}
{The Agency for Digital Italy}.
\newblock {Italian Observatory on Artificial Intelligence }.
\newblock \url{{https://ia.italia.it/en/ai-observatory/}},  2020.
\newblock Online; accessed 25-April-2020.

\bibitem[Krausov{\'a}(2020)]{krausova2020czech}
Krausov{\'a}, A.
\newblock {Czech Republic's AI Observatory and Forum}.
\newblock {\em The Lawyer Quarterly} {\bf 2020}, {\em 1}.

\bibitem[{Denkfabrik}(2020)]{germanAI}
{Denkfabrik}.
\newblock {AI Observatory}.
\newblock \url{{https://www.denkfabrik-bmas.de/en/projects/ai-observatory}},
  2020.
\newblock {Digitale Arbeitsgesellschaft}; accessed 28-November-2020.

\bibitem[{OECD.AI}(2020)]{oecdAI}
{OECD.AI}.
\newblock {OECD AI Policy Observatory }.
\newblock \url{{https://oecd.ai/}},  2020.
\newblock Online; accessed 25-April-2020.

\bibitem[Yampolskiy(2019)]{yampolskiy2019predicting}
Yampolskiy, R.V.
\newblock {Predicting future AI failures from historic examples}.
\newblock {\em foresight} {\bf 2019}.

\bibitem[McGregor(2020)]{mcgregor2020preventing}
McGregor, S.
\newblock {Preventing Repeated Real World AI Failures by Cataloging Incidents:
  The AI Incident Database}.
\newblock {\em arXiv preprint arXiv:2011.08512} {\bf 2020}.

\bibitem[Aliman(2020)]{aliman2020hybrid}
Aliman, N.M.
\newblock {Hybrid Cognitive-Affective Strategies for AI Safety}.
\newblock PhD thesis, Utrecht University,  2020.

\bibitem[Woo(2019)]{woo2019downward}
Woo, G.
\newblock {Downward Counterfactual Search for Extreme Events}.
\newblock {\em Frontiers in Earth Science} {\bf 2019}, {\em 7},~340.

\bibitem[Roese and Epstude(2017)]{roese2017functional}
Roese, N.J.; Epstude, K.
\newblock The functional theory of counterfactual thinking: New evidence, new
  challenges, new insights. In {\em Advances in experimental social
  psychology}; Elsevier,  2017; Vol.~56, pp. 1--79.

\bibitem[Aliman \em{et~al.}(2020)Aliman, Elands, H{\"u}rst, Kester,
  Th{\'o}risson, Werkhoven, Yampolskiy, and Ziesche]{aliman2020error}
Aliman, N.M.; Elands, P.; H{\"u}rst, W.; Kester, L.; Th{\'o}risson, K.R.;
  Werkhoven, P.; Yampolskiy, R.; Ziesche, S.
\newblock {Error-Correction for AI Safety}.
\newblock  International Conference on Artificial General Intelligence.
  Springer,  2020, pp. 12--22.

\bibitem[Brundage \em{et~al.}(2018)Brundage, Avin, Clark, Toner, Eckersley,
  Garfinkel, Dafoe, Scharre, Zeitzoff, Filar, et~al.]{brundage2018malicious}
Brundage, M.; Avin, S.; Clark, J.; Toner, H.; Eckersley, P.; Garfinkel, B.;
  Dafoe, A.; Scharre, P.; Zeitzoff, T.; Filar, B.; others.
\newblock {The malicious use of artificial intelligence: Forecasting,
  prevention, and mitigation}.
\newblock {\em arXiv preprint arXiv:1802.07228} {\bf 2018}.

\bibitem[Pistono and Yampolskiy(2016)]{DBLP:journals/corr/PistonoY16}
Pistono, F.; Yampolskiy, R.V.
\newblock {Unethical Research: How to Create a Malevolent Artificial
  Intelligence}.
\newblock {\em CoRR} {\bf 2016}, {\em abs/1605.02817},
  \href{http://xxx.lanl.gov/abs/1605.02817}{{\normalfont [1605.02817]}}.

\bibitem[Aliman \em{et~al.}(2020)Aliman, Kester, Werkhoven, and
  Ziesche]{Delphi2}
Aliman, N.M.; Kester, L.; Werkhoven, P.; Ziesche, S.
\newblock {Sustainable AI Safety?}
\newblock {\em Delphi -- Interdisciplinary review of emerging technologies}
  {\bf 2020}, {\em 2},~226--233.

\bibitem[Aliman \em{et~al.}(2019)Aliman, Kester, Werkhoven, and
  Yampolskiy]{aliman2019orthogonality}
Aliman, N.M.; Kester, L.; Werkhoven, P.; Yampolskiy, R.
\newblock Orthogonality-based disentanglement of responsibilities for ethical
  intelligent systems.
\newblock  International Conference on Artificial General Intelligence.
  Springer,  2019, pp. 22--31.

\bibitem[Cancila \em{et~al.}(2018)Cancila, Gerstenmayer, Espinoza, and
  Passerone]{cancila2018sharpening}
Cancila, D.; Gerstenmayer, J.L.; Espinoza, H.; Passerone, R.
\newblock {Sharpening the scythe of technological change: Socio-technical
  challenges of autonomous and adaptive cyber-physical systems}.
\newblock {\em Designs} {\bf 2018}, {\em 2},~52.

\bibitem[Martin~Jr \em{et~al.}(2020)Martin~Jr, Prabhakaran, Kuhlberg, Smart,
  and Isaac]{martin2020extending}
Martin~Jr, D.; Prabhakaran, V.; Kuhlberg, J.; Smart, A.; Isaac, W.S.
\newblock {Extending the Machine Learning Abstraction Boundary: A Complex
  Systems Approach to Incorporate Societal Context}.
\newblock {\em arXiv preprint arXiv:2006.09663} {\bf 2020}.

\bibitem[Scott and Yampolskiy(2020)]{scott2020classification}
Scott, P.J.; Yampolskiy, R.V.
\newblock {Classification Schemas for Artificial Intelligence Failures}.
\newblock {\em Delphi-Interdisciplinary Review of Emerging Technologies} {\bf
  2020}, {\em 2},~186--199.

\bibitem[Gray \em{et~al.}(2012)Gray, Waytz, and Young]{gray2012moral}
Gray, K.; Waytz, A.; Young, L.
\newblock {The moral dyad: A fundamental template unifying moral judgment}.
\newblock {\em Psychological Inquiry} {\bf 2012}, {\em 23},~206--215.

\bibitem[Schein and Gray(2018)]{schein2018theory}
Schein, C.; Gray, K.
\newblock {The theory of dyadic morality: Reinventing moral judgment by
  redefining harm}.
\newblock {\em Personality and Social Psychology Review} {\bf 2018}, {\em
  22},~32--70.

\bibitem[Gray \em{et~al.}(2017)Gray, Schein, and Cameron]{gray2017think}
Gray, K.; Schein, C.; Cameron, C.D.
\newblock {How to think about emotion and morality: Circles, not arrows}.
\newblock {\em Current opinion in psychology} {\bf 2017}, {\em 17},~41--46.

\bibitem[Popper(1966)]{popper1966poverty}
Popper, K.R.
\newblock {\em The poverty of historicism}; Routledge \& Kegan Paul,  1966.

\bibitem[Aliman and Kester(2020)]{aliman2020mal}
Aliman, N.; Kester, L.
\newblock Malicious Design in AIVR, Falsehood and Cybersecurity-oriented
  Immersive Defenses.
\newblock  2020 IEEE International Conference on Artificial Intelligence and
  Virtual Reality (AIVR). IEEE,  2020, p. to appear.

\bibitem[Harwell(2019)]{harwell}
Harwell, D.
\newblock {An artificial-intelligence first: Voice-mimicking software
  reportedly used in a major theft}.
\newblock
  \url{https://www.washingtonpost.com/technology/2019/09/04/an-artificial-intelligence-first-voice-mimicking-software-reportedly-used-major-theft/},
   2019.
\newblock {The Washington Post}; accessed 04-August-2020.

\bibitem[Rohrlich(2020)]{rohrlich}
Rohrlich, J.
\newblock {Romance Scammer Used Deepfakes to Impersonate a Navy Admiral and
  Bilk Widow Out of Nearly \$300,000}.
\newblock
  \url{https://www.thedailybeast.com/romance-scammer-used-deepfakes-to-impersonate-a-navy-admiral-and-bilk-widow-out-of-nearly-dollar300000},
   2020.
\newblock {Daily Beastl}; accessed 08-November-2020.

\bibitem[Rushing(2020)]{rushing}
Rushing, E.
\newblock {A Philly lawyer nearly wired \$9,000 to a stranger impersonating his
  son’s voice, showing just how smart scammers are getting}.
\newblock
  \url{https://www.inquirer.com/news/voice-scam-impersonation-fraud-bail-bond-artificial-intelligence-20200309.html},
   2020.
\newblock {The Philadelphia Inquirer}; accessed 04-August-2020.

\bibitem[Stupp(2019)]{stupp}
Stupp, C.
\newblock {Fraudsters Used AI to Mimic CEO's Voice in Unusual Cybercrime Case}.
\newblock
  \url{https://www.wsj.com/articles/fraudsters-use-ai-to-mimic-ceos-voice-in-unusual-cybercrime-case-11567157402},
   2019.
\newblock {The Wall Street Journal}; accessed 04-August-2020.

\bibitem[Gieseke(2020)]{gieseke2020new}
Gieseke, A.P.
\newblock {" The New Weapon of Choice": Law's Current Inability to Properly
  Address Deepfake Pornography}.
\newblock {\em Vanderbilt Law Review} {\bf 2020}, {\em 73},~1479--1515.

\bibitem[Ajder \em{et~al.}(2019)Ajder, Patrini, Cavalli, and
  Cullen]{ajder2019state}
Ajder, H.; Patrini, G.; Cavalli, F.; Cullen, L.
\newblock {The State of Deepfakes: Landscape, Threats, and Impact}.
\newblock {\em Amsterdam: Deeptrace} {\bf 2019}.

\bibitem[Alba(2019)]{alba}
Alba, D.
\newblock {Facebook Discovers Fakes That Show Evolution of Disinformation}.
\newblock
  \url{https://www.nytimes.com/2019/12/20/business/facebook-ai-generated-profiles.html},
   2019.
\newblock {The New York Times}; accessed 04-August-2020.

\bibitem[{Reuters}(2020)]{reut}
{Reuters}.
\newblock {Deepfake Used to Attack Activist Couple Shows New Disinformation
  Frontier}.
\newblock
  \url{https://gadgets.ndtv.com/internet/features/deepfake-oliver-taylor-mazen-masri-terrorist-accuse-london-university-of-birmingham-student-fake-profile-22640449},
   2020.
\newblock {Reuters}; accessed 08-November-2020.

\bibitem[Cole and Maiberg(2019)]{cole}
Cole, S.; Maiberg, E.
\newblock {Deepfake Porn Is Evolving to Give People Total Control Over Women's
  Bodies}.
\newblock
  \url{https://www.vice.com/en/article/9keen8/deepfake-porn-is-evolving-to-give-people-total-control-over-womens-bodies},
   2019.
\newblock {VICE}; accessed 08-November-2020.

\bibitem[Hao(2020)]{Hao}
Hao, K.
\newblock {A deepfake bot is being used to "undress" underage girls}.
\newblock
  \url{https://www.technologyreview.com/2020/10/20/1010789/ai-deepfake-bot-undresses-women-and-underage-girls/},
   2020.
\newblock {MIT Technology Review}; accessed 08-November-2020.

\bibitem[Corera(2020)]{corera}
Corera, G.
\newblock {UK spies will need artificial intelligence - Rusi report}.
\newblock \url{https://www.bbc.com/news/technology-52415775},  2020.
\newblock {BBC}; accessed 08-November-2020.

\bibitem[Satter(2019)]{satter}
Satter, R.
\newblock {Experts: Spy used AI-generated face to connect with targets}.
\newblock \url{https://apnews.com/article/bc2f19097a4c4fffaa00de6770b8a60d},
  2019.
\newblock {AP News}; accessed 04-August-2020.

\bibitem[Probyn and Doran(2020)]{probyn}
Probyn, A.; Doran, M.
\newblock {One Month, 500,000 Face Scans: How China Is Using A.I. to Profile a
  Minority}.
\newblock
  \url{https://www.abc.net.au/news/2020-09-14/chinese-data-leak-linked-to-military-names-australians/12656668},
   2020.
\newblock {ABC News}; accessed 04-August-2020.

\bibitem[Mozur(2019)]{mozur}
Mozur, P.
\newblock {China's 'hybrid war': Beijing's mass surveillance of Australia and
  the world for secrets and scandal}.
\newblock
  \url{https://www.nytimes.com/2019/04/14/technology/china-surveillance-artificial-intelligence-racial-profiling.html},
   2019.
\newblock {The New York Times}; accessed 04-August-2020.

\bibitem[Neekhara \em{et~al.}(2020)Neekhara, Hussain, Jere, Koushanfar, and
  McAuley]{neekhara2020adversarial}
Neekhara, P.; Hussain, S.; Jere, M.; Koushanfar, F.; McAuley, J.
\newblock {Adversarial Deepfakes: Evaluating Vulnerability of Deepfake
  Detectors to Adversarial Examples}.
\newblock {\em arXiv preprint arXiv:2002.12749} {\bf 2020}.

\bibitem[Zang \em{et~al.}(2020)Zang, Sweeney, and Weiss]{zangt}
Zang, J.; Sweeney, L.; Weiss, M.
\newblock {The real threat of fake voices in a time of crisis}.
\newblock
  \url{https://techcrunch.com/2020/05/16/the-real-threat-of-fake-voices-in-a-time-of-crisis/?guccounter=1},
   2020.
\newblock {Techcrunch}; accessed 08-November-2020.

\bibitem[O'Donnell(2020)]{donnell}
O'Donnell, L.
\newblock {Black Hat 2020: Open-Source AI to Spur Wave of 'Synthetic Media'
  Attacks}.
\newblock
  \url{https://threatpost.com/black-hat-2020-open-source-ai-to-spur-wave-of-synthetic-media-attacks/158066/},
   2020.
\newblock {threatpost}; accessed 08-November-2020.

\bibitem[Jr \em{et~al.}(2020)Jr, Note, Spellchecker, and
  Yampolskiy]{TransformerJrManuscript-GPTWSC}
Jr, G.P.T.; Note, E.X.; Spellchecker, M.S.; Yampolskiy, R.
\newblock {When Should Co-Authorship Be Given to AI?}
\newblock \url{https://philarchive.org/archive/GPTWSCv1 },  2020.
\newblock {Unpublished, PhilArchive}; accessed 08-November-2020.

\bibitem[Zhang \em{et~al.}(2003)Zhang, Zhou, Qin, and Liu]{zhang2003honeypot}
Zhang, F.; Zhou, S.; Qin, Z.; Liu, J.
\newblock Honeypot: a supplemented active defense system for network security.
\newblock  Proceedings of the Fourth International Conference on Parallel and
  Distributed Computing, Applications and Technologies. IEEE,  2003, pp.
  231--235.

\bibitem[Nelson \em{et~al.}(2020)Nelson, Esq., and Simek]{nelson}
Nelson, S.D.; Esq..; Simek, J.W.
\newblock {Video and Audio Deepfakes: What Lawyers Need to Know}.
\newblock
  \url{https://www.masslomap.org/video-and-audio-deepfakes-what-lawyers-need-to-know-guest-post/},
   2020.
\newblock {Sensei Enterprises, Inc.}; accessed 08-November-2020.

\bibitem[Chen \em{et~al.}(2019)Chen, Liu, Xiang, Niu, Tong, and
  Han]{chen2019adversarial}
Chen, T.; Liu, J.; Xiang, Y.; Niu, W.; Tong, E.; Han, Z.
\newblock Adversarial attack and defense in reinforcement learning-from AI
  security view.
\newblock {\em Cybersecurity} {\bf 2019}, {\em 2},~11.

\bibitem[Spocchia(2020)]{spocchia}
Spocchia, G.
\newblock {Republican candidate shares conspiracy theory that George Floyd
  murder was faked}.
\newblock
  \url{https://www.independent.co.uk/news/world/americas/us-politics/george-floyd-murder-fake-conspiracy-theory-hoax-republican-gop-missouri-a9580896.html},
   2020.
\newblock {Independent}; accessed 04-August-2020.

\bibitem[Hao(2019)]{Hao2}
Hao, K.
\newblock {The Biggest Threat of Deepfakes Isn’t the Deepfakes Themselves}.
\newblock
  \url{https://www.technologyreview.com/2019/10/10/132667/the-biggest-threat-of-deepfakes-isnt-the-deepfakes-themselves/},
   2019.
\newblock {MIT Technology Review}; accessed 08-November-2020.

\bibitem[Bilge and Dumitra{\c{s}}(2012)]{bilge2012before}
Bilge, L.; Dumitra{\c{s}}, T.
\newblock {Before we knew it: an empirical study of zero-day attacks in the
  real world}.
\newblock  Proceedings of the 2012 ACM conference on Computer and
  communications security,  2012, pp. 833--844.

\bibitem[Carlini and Wagner(2017)]{carlini2017adversarial}
Carlini, N.; Wagner, D.
\newblock {Adversarial examples are not easily detected: Bypassing ten
  detection methods}.
\newblock  Proceedings of the 10th ACM Workshop on Artificial Intelligence and
  Security,  2017, pp. 3--14.

\bibitem[Carlini(2020)]{carlini2020partial}
Carlini, N.
\newblock {A Partial Break of the Honeypots Defense to Catch Adversarial
  Attacks},  2020,  \href{http://xxx.lanl.gov/abs/2009.10975}{{\normalfont
  [arXiv:cs.CR/2009.10975]}}.

\bibitem[Papernot \em{et~al.}(2016)Papernot, McDaniel, Sinha, and
  Wellman]{papernot2016towards}
Papernot, N.; McDaniel, P.; Sinha, A.; Wellman, M.
\newblock Towards the science of security and privacy in machine learning.
\newblock {\em arXiv preprint arXiv:1611.03814} {\bf 2016}.

\bibitem[Tramer \em{et~al.}(2020)Tramer, Carlini, Brendel, and
  Madry]{tramer2020adaptive}
Tramer, F.; Carlini, N.; Brendel, W.; Madry, A.
\newblock On adaptive attacks to adversarial example defenses.
\newblock {\em arXiv preprint arXiv:2002.08347} {\bf 2020}.

\bibitem[Kirat \em{et~al.}(2018)Kirat, Jang, and
  Stoecklin]{kirat2018deeplocker}
Kirat, D.; Jang, J.; Stoecklin, M.
\newblock {Deeplocker--Concealing Targeted Attacks with AI Locksmithing}.
\newblock {\em Blackhat USA} {\bf 2018}.

\bibitem[Qiu \em{et~al.}(2019)Qiu, Xiao, Yang, Yan, Lee, and
  Li]{qiu2019semanticadv}
Qiu, H.; Xiao, C.; Yang, L.; Yan, X.; Lee, H.; Li, B.
\newblock {Semanticadv: Generating adversarial examples via
  attribute-conditional image editing}.
\newblock {\em arXiv preprint arXiv:1906.07927} {\bf 2019}.

\bibitem[Carlini and Farid(2020)]{carlini2020evading}
Carlini, N.; Farid, H.
\newblock {Evading Deepfake-Image Detectors with White-and Black-Box Attacks}.
\newblock  Proceedings of the IEEE/CVF Conference on Computer Vision and
  Pattern Recognition Workshops,  2020, pp. 658--659.

\bibitem[Xu \em{et~al.}(2020)Xu, Zhang, Liu, Fan, Sun, Chen, Chen, Wang, and
  Lin]{xu2020adversarial}
Xu, K.; Zhang, G.; Liu, S.; Fan, Q.; Sun, M.; Chen, H.; Chen, P.Y.; Wang, Y.;
  Lin, X.
\newblock Adversarial t-shirt! evading person detectors in a physical world.
\newblock  European Conference on Computer Vision. Springer,  2020, pp.
  665--681.

\bibitem[Wallace \em{et~al.}(2019)Wallace, Feng, Kandpal, Gardner, and
  Singh]{wallace2019universal}
Wallace, E.; Feng, S.; Kandpal, N.; Gardner, M.; Singh, S.
\newblock {Universal Adversarial Triggers for Attacking and Analyzing NLP}.
\newblock  Proceedings of the 2019 Conference on Empirical Methods in Natural
  Language Processing and the 9th International Joint Conference on Natural
  Language Processing (EMNLP-IJCNLP),  2019.

\bibitem[Cheng \em{et~al.}(2020)Cheng, Juefei-Xu, Guo, Fu, Xie, Lin, Lin, and
  Liu]{cheng2020adversarial}
Cheng, Y.; Juefei-Xu, F.; Guo, Q.; Fu, H.; Xie, X.; Lin, S.W.; Lin, W.; Liu, Y.
\newblock {Adversarial Exposure Attack on Diabetic Retinopathy Imagery}.
\newblock {\em arXiv preprint arXiv:2009.09231} {\bf 2020}.

\bibitem[Finlayson \em{et~al.}(2019)Finlayson, Bowers, Ito, Zittrain, Beam, and
  Kohane]{finlayson2019adversarial}
Finlayson, S.G.; Bowers, J.D.; Ito, J.; Zittrain, J.L.; Beam, A.L.; Kohane,
  I.S.
\newblock Adversarial attacks on medical machine learning.
\newblock {\em Science} {\bf 2019}, {\em 363},~1287--1289.

\bibitem[Han \em{et~al.}(2020)Han, Hu, Foschini, Chinitz, Jankelson, and
  Ranganath]{han2020deep}
Han, X.; Hu, Y.; Foschini, L.; Chinitz, L.; Jankelson, L.; Ranganath, R.
\newblock Deep learning models for electrocardiograms are susceptible to
  adversarial attack.
\newblock {\em Nature Medicine} {\bf 2020}, pp. 1--4.

\bibitem[Zhang \em{et~al.}(2020)Zhang, Wu, Ding, Luo, Lin, Jung, and
  Chavarriaga]{zhang2020tiny}
Zhang, X.; Wu, D.; Ding, L.; Luo, H.; Lin, C.T.; Jung, T.P.; Chavarriaga, R.
\newblock Tiny noise, big mistakes: adversarial perturbations induce errors in
  brain-computer interface spellers.
\newblock {\em National Science Review} {\bf 2020}.

\bibitem[Zhou \em{et~al.}(2018)Zhou, Tang, Wang, Han, Liu, and
  Zhang]{zhou2018invisible}
Zhou, Z.; Tang, D.; Wang, X.; Han, W.; Liu, X.; Zhang, K.
\newblock {Invisible mask: Practical attacks on face recognition with
  infrared}.
\newblock {\em arXiv preprint arXiv:1803.04683} {\bf 2018}.

\bibitem[Cao \em{et~al.}(2019)Cao, Xiao, Cyr, Zhou, Park, Rampazzi, Chen, Fu,
  and Mao]{cao2019adversarial}
Cao, Y.; Xiao, C.; Cyr, B.; Zhou, Y.; Park, W.; Rampazzi, S.; Chen, Q.A.; Fu,
  K.; Mao, Z.M.
\newblock {Adversarial sensor attack on LiDAR-based perception in autonomous
  driving}.
\newblock  Proceedings of the 2019 ACM SIGSAC Conference on Computer and
  Communications Security,  2019, pp. 2267--2281.

\bibitem[Povolny and Trivedi(2020)]{mcafee}
Povolny, S.; Trivedi, S.
\newblock {Model Hacking ADAS to Pave Safer Roads for Autonomous Vehicles}.
\newblock
  \url{{https://www.mcafee.com/blogs/other-blogs/mcafee-labs/model-hacking-adas-to-pave-safer-roads-for-autonomous-vehicles/}},
   2020.
\newblock McAfee; accessed 08-November-2020.

\bibitem[Chen \em{et~al.}(2020)Chen, Yuan, Zhang, Zhao, Zhang, Chen, and
  Wang]{247642}
Chen, Y.; Yuan, X.; Zhang, J.; Zhao, Y.; Zhang, S.; Chen, K.; Wang, X.
\newblock {Devil{\textquoteright}s Whisper: A General Approach for Physical
  Adversarial Attacks against Commercial Black-box Speech Recognition Devices}.
\newblock  29th {USENIX} Security Symposium ({USENIX} Security 20). {USENIX}
  Association,  2020, pp. 2667--2684.

\bibitem[Li \em{et~al.}(2019)Li, Qu, Li, Szurley, Kolter, and
  Metze]{li2019adversarial}
Li, J.; Qu, S.; Li, X.; Szurley, J.; Kolter, J.Z.; Metze, F.
\newblock {Adversarial music: Real world audio adversary against wake-word
  detection system}.
\newblock  Advances in Neural Information Processing Systems,  2019, pp.
  11931--11941.

\bibitem[Wu \em{et~al.}(2020)Wu, Zhou, Liu, Liu, and Zhu]{wu2020decision}
Wu, J.; Zhou, M.; Liu, S.; Liu, Y.; Zhu, C.
\newblock {Decision-based Universal Adversarial Attack}.
\newblock {\em arXiv preprint arXiv:2009.07024} {\bf 2020}.

\bibitem[Shumailov \em{et~al.}(2020)Shumailov, Zhao, Bates, Papernot, Mullins,
  and Anderson]{shumailov2020sponge}
Shumailov, I.; Zhao, Y.; Bates, D.; Papernot, N.; Mullins, R.; Anderson, R.
\newblock {Sponge Examples: Energy-Latency Attacks on Neural Networks}.
\newblock {\em arXiv preprint arXiv:2006.03463} {\bf 2020}.

\bibitem[Cin{\`a} \em{et~al.}(2020)Cin{\`a}, Torcinovich, and
  Pelillo]{cina2020black}
Cin{\`a}, A.E.; Torcinovich, A.; Pelillo, M.
\newblock {A Black-box Adversarial Attack for Poisoning Clustering}.
\newblock {\em arXiv preprint arXiv:2009.05474} {\bf 2020}.

\bibitem[Chitpin(2013)]{chitpin2013should}
Chitpin, S.
\newblock Should Popper’s view of rationality be used for promoting teacher
  knowledge?
\newblock {\em Educational Philosophy and Theory} {\bf 2013}, {\em
  45},~833--844.

\bibitem[Obermeyer \em{et~al.}(2019)Obermeyer, Powers, Vogeli, and
  Mullainathan]{obermeyer2019dissecting}
Obermeyer, Z.; Powers, B.; Vogeli, C.; Mullainathan, S.
\newblock Dissecting racial bias in an algorithm used to manage the health of
  populations.
\newblock {\em Science} {\bf 2019}, {\em 366},~447--453.

\bibitem[Hill(2020)]{hill2020wrongfully}
Hill, K.
\newblock Wrongfully accused by an algorithm.
\newblock {\em The New York Times} {\bf 2020}.

\bibitem[Buolamwini and Gebru(2018)]{buolamwini2018gender}
Buolamwini, J.; Gebru, T.
\newblock Gender shades: Intersectional accuracy disparities in commercial
  gender classification.
\newblock  Conference on fairness, accountability and transparency,  2018, pp.
  77--91.

\bibitem[Da~Costa(2020)]{joy}
Da~Costa, C.
\newblock {The Women Geniuses Taking on Racial and Gender Bias in AI -- and
  Amazon }.
\newblock
  \url{https://www.thedailybeast.com/the-women-geniuses-taking-on-racial-and-gender-bias-in-artificial-intelligence-and-amazon},
   2020.
\newblock {The Daily Beast}; accessed 23-May-2020.

\bibitem[Larrazabal \em{et~al.}(2020)Larrazabal, Nieto, Peterson, Milone, and
  Ferrante]{larrazabal2020gender}
Larrazabal, A.J.; Nieto, N.; Peterson, V.; Milone, D.H.; Ferrante, E.
\newblock Gender imbalance in medical imaging datasets produces biased
  classifiers for computer-aided diagnosis.
\newblock {\em Proceedings of the National Academy of Sciences} {\bf 2020},
  {\em 117},~12592--12594.

\bibitem[Prabhu and Birhane(2020)]{prabhu2020large}
Prabhu, V.U.; Birhane, A.
\newblock {Large image datasets: A pyrrhic win for computer vision?}
\newblock {\em arXiv preprint arXiv:2006.16923} {\bf 2020}.

\bibitem[Jain \em{et~al.}(2020)Jain, Olmo, Sengupta, Manikonda, and
  Kambhampati]{jain2020imperfect}
Jain, N.; Olmo, A.; Sengupta, S.; Manikonda, L.; Kambhampati, S.
\newblock {Imperfect imaganation: Implications of gans exacerbating biases on
  facial data augmentation and snapchat selfie lenses}.
\newblock {\em arXiv preprint arXiv:2001.09528} {\bf 2020}.

\bibitem[Kempsell(2020)]{kempsell}
Kempsell, R.
\newblock {Ofqual pauses study into whether AI could be used to mark exams}.
\newblock
  \url{{https://www.thetimes.co.uk/article/robot-exam-marking-project-is-put-on-hold-vvrm753l3}},
   2020.
\newblock {The Times}; accessed 10-November-2020.

\bibitem[Huchel(2020)]{huchel}
Huchel, B.
\newblock {Artificial intelligence examines best ways to keep parolees from
  recommitting crimes}.
\newblock
  \url{https://phys.org/news/2020-08-artificial-intelligence-ways-parolees-recommitting.html},
   2020.
\newblock {Phys Org}; accessed 20-August-2020.

\bibitem[Cushing(2020)]{cushing}
Cushing, T.
\newblock {Harrisburg University Researchers Claim Their 'Unbiased' Facial
  Recognition Software Can Identify Potential Criminals}.
\newblock
  \url{https://www.techdirt.com/articles/20200505/17090244442/harrisburg-university-researchers-claim-their-unbiased-facial-recognition-software-can-identify-potential-criminals.shtml},
   2020.
\newblock {techdirt}; accessed 02-November-2020.

\bibitem[{Harrisburg University }(2020)]{facial2}
{Harrisburg University }.
\newblock {HU facial recognition software predicts criminality}.
\newblock \url{http://archive.is/N1HVe#selection-1509.0-1509.51},  2020.
\newblock Online; accessed 23-May-2020.

\bibitem[Pascu(2020)]{facial3}
Pascu, L.
\newblock {Biometric software that allegedly predicts criminals based on their
  face sparks industry controversy}.
\newblock
  \url{https://www.biometricupdate.com/202005/biometric-software-that-allegedly-predicts-criminals-based-on-their-face-sparks-industry-controversy},
   2020.
\newblock Biometric; accessed 23-May-2020.

\bibitem[Barrett \em{et~al.}(2019)Barrett, Adolphs, Marsella, Martinez, and
  Pollak]{barrett2019emotional}
Barrett, L.F.; Adolphs, R.; Marsella, S.; Martinez, A.M.; Pollak, S.D.
\newblock Emotional expressions reconsidered: challenges to inferring emotion
  from human facial movements.
\newblock {\em Psychological Science in the Public Interest} {\bf 2019}, {\em
  20},~1--68.

\bibitem[Gendron \em{et~al.}(2020)Gendron, Hoemann, Crittenden, Mangola, Ruark,
  and Barrett]{gendron2020emotion}
Gendron, M.; Hoemann, K.; Crittenden, A.N.; Mangola, S.M.; Ruark, G.A.;
  Barrett, L.F.
\newblock {Emotion perception in Hadza Hunter-Gatherers}.
\newblock {\em Scientific reports} {\bf 2020}, {\em 10},~1--17.

\bibitem[Crawford \em{et~al.}(2019)Crawford, Dobbe, Dryer, Fried, Green,
  Kaziunas, Kak, Mathur, McElroy, S{\'a}nchez, et~al.]{now}
Crawford, K.; Dobbe, R.; Dryer, T.; Fried, G.; Green, B.; Kaziunas, E.; Kak,
  A.; Mathur, V.; McElroy, E.; S{\'a}nchez, A.N.; others.
\newblock {AI Now 2019 Report}.
\newblock \url{https://ainowinstitute.org/AI_Now_2019_Report.pdf},  2019.
\newblock AI Now Institute; accessed 23-May-2020.

\bibitem[Lieber(2018)]{lieber}
Lieber, C.
\newblock {Tech companies use "persuasive design" to get us hooked.
  Psychologists say it’s unethical.}
\newblock
  \url{{https://www.vox.com/2018/8/8/17664580/persuasive-technology-psychology}},
   2018.
\newblock Vox; accessed 08-November-2020.

\bibitem[Jakubowski(2019)]{jakubowski2019s}
Jakubowski, G.
\newblock {What’s not to like? Social media as information operations force
  multiplier}.
\newblock {\em Joint Force Quarterly} {\bf 2019}, {\em 3},~8--17.

\bibitem[Sawers(2020)]{sawers}
Sawers, P.
\newblock {The Social Dilemma: How digital platforms pose an existential threat
  to society}.
\newblock
  \url{https://venturebeat.com/2020/09/02/the-social-dilemma-how-digital-platforms-pose-an-existential-threat-to-society/},
   2020.
\newblock {VentureBeat}; accessed 02-November-2020.

\bibitem[Chikhale and Gohad(2018)]{chikhale2018multidimensional}
Chikhale, S.; Gohad, V.
\newblock {Multidimensional Construct About The Robot Citizenship Law's In
  Saudi Arabia}.
\newblock {\em International Journal of Innovative Research and Advanced
  Studies (IJIRAS)} {\bf 2018}, {\em 5},~106--108.

\bibitem[Yam \em{et~al.}(2020)Yam, Bigman, Tang, Ilies, De~Cremer, Soh, and
  Gray]{yam2020robots}
Yam, K.C.; Bigman, Y.E.; Tang, P.M.; Ilies, R.; De~Cremer, D.; Soh, H.; Gray,
  K.
\newblock {Robots at work: People prefer—and forgive—service robots with
  perceived feelings.}
\newblock {\em Journal of Applied Psychology} {\bf 2020}.

\bibitem[Orabi \em{et~al.}(2020)Orabi, Mouheb, Al~Aghbari, and
  Kamel]{orabi2020detection}
Orabi, M.; Mouheb, D.; Al~Aghbari, Z.; Kamel, I.
\newblock {Detection of Bots in Social Media: A Systematic Review}.
\newblock {\em Information Processing \& Management} {\bf 2020}, {\em
  57},~102250.

\bibitem[Prier(2017)]{prier2017commanding}
Prier, J.
\newblock {Commanding the trend: Social media as information warfare}.
\newblock {\em Strategic Studies Quarterly} {\bf 2017}, {\em 11},~50--85.

\bibitem[Letter(2018)]{apapsy}
Letter, O.
\newblock {Our letter to the APA}.
\newblock \url{https://screentimenetwork.org/apa},  2018.
\newblock {Online}; accessed 02-November-2020.

\bibitem[Theriault \em{et~al.}(2020)Theriault, Young, and
  Barrett]{theriault2020sense}
Theriault, J.E.; Young, L.; Barrett, L.F.
\newblock {The sense of should: A biologically-based framework for modeling
  social pressure}.
\newblock {\em Physics of Life Reviews} {\bf 2020}.

\bibitem[Anderson and Jiang(2018)]{anderson2018teens}
Anderson, M.; Jiang, J.
\newblock Teens’ social media habits and experiences.
\newblock {\em Pew Research Center} {\bf 2018}, {\em 28}.

\bibitem[Barber{\'a} and Zeitzoff(2018)]{barbera2018new}
Barber{\'a}, P.; Zeitzoff, T.
\newblock The new public address system: why do world leaders adopt social
  media?
\newblock {\em International Studies Quarterly} {\bf 2018}, {\em 62},~121--130.

\bibitem[Franchina and Coco(2018)]{franchina2018influence}
Franchina, V.; Coco, G.L.
\newblock The influence of social media use on body image concerns.
\newblock {\em International Journal of Psychoanalysis and Education} {\bf
  2018}, {\em 10},~5--14.

\bibitem[Halfmann and Rieger(2019)]{halfmann2019permanently}
Halfmann, A.; Rieger, D.
\newblock {Permanently on call: The effects of social pressure on smartphone
  users’ self-control, need satisfaction, and well-being}.
\newblock {\em Journal of Computer-Mediated Communication} {\bf 2019}, {\em
  24},~165--181.

\bibitem[Stieger and Lewetz(2018)]{stieger2018week}
Stieger, S.; Lewetz, D.
\newblock {A week without using social media: Results from an ecological
  momentary intervention study using smartphones}.
\newblock {\em Cyberpsychology, Behavior, and Social Networking} {\bf 2018},
  {\em 21},~618--624.

\bibitem[Ferrara and Yang(2015)]{ferrara2015measuring}
Ferrara, E.; Yang, Z.
\newblock Measuring emotional contagion in social media.
\newblock {\em PloS one} {\bf 2015}, {\em 10},~e0142390.

\bibitem[Luxton \em{et~al.}(2012)Luxton, June, and Fairall]{luxton2012social}
Luxton, D.D.; June, J.D.; Fairall, J.M.
\newblock Social media and suicide: a public health perspective.
\newblock {\em American journal of public health} {\bf 2012}, {\em
  102},~S195--S200.

\bibitem[Lane(2020)]{lane2020nist}
Lane, L.
\newblock {NIST finds flaws in facial checks on people with Covid masks}.
\newblock {\em Biometric Technology Today} {\bf 2020}.

\bibitem[Mundial \em{et~al.}(2020)Mundial, Hassan, Tiwana, Qureshi, and
  Alanazi]{mundial2020towards}
Mundial, I.Q.; Hassan, M.S.U.; Tiwana, M.I.; Qureshi, W.S.; Alanazi, E.
\newblock {Towards Facial Recognition Problem in COVID-19 Pandemic}.
\newblock  2020 4rd International Conference on Electrical, Telecommunication
  and Computer Engineering (ELTICOM). IEEE,  2020, pp. 210--214.

\bibitem[Ngan \em{et~al.}(2020)Ngan, Grother, and Hanaoka]{ngan2020ongoing}
Ngan, M.L.; Grother, P.J.; Hanaoka, K.K.
\newblock {Ongoing Face Recognition Vendor Test (FRVT) Part 6A: Face
  recognition accuracy with masks using pre-COVID-19 algorithms}.
\newblock {\em National Institute of Standards and Technology} {\bf 2020}.

\bibitem[Krishna \em{et~al.}(2019)Krishna, Tomar, Parikh, Papernot, and
  Iyyer]{krishna2019thieves}
Krishna, K.; Tomar, G.S.; Parikh, A.P.; Papernot, N.; Iyyer, M.
\newblock {Thieves on Sesame Street! Model Extraction of BERT-based APIs}.
\newblock {\em arXiv preprint arXiv:1910.12366} {\bf 2019}.

\bibitem[Taylor(2020)]{taylorj}
Taylor, J.
\newblock {Facebook incorrectly removes picture of Aboriginal men in chains
  because of 'nudity' }.
\newblock
  \url{https://www.theguardian.com/technology/2020/jun/13/facebook-incorrectly-removes-picture-of-aboriginal-men-in-chains-because-of-nudity},
   2020.
\newblock {The Guardian}; accessed 02-November-2020.

\bibitem[DeCamp and Lindvall(2020)]{decamp2020latent}
DeCamp, M.; Lindvall, C.
\newblock Latent bias and the implementation of artificial intelligence in
  medicine.
\newblock {\em Journal of the American Medical Informatics Association} {\bf
  2020}.

\bibitem[Kaushal \em{et~al.}(2020)Kaushal, Altman, and
  Langlotz]{kaushal2020geographic}
Kaushal, A.; Altman, R.; Langlotz, C.
\newblock {Geographic Distribution of US Cohorts Used to Train Deep Learning
  Algorithms}.
\newblock {\em Jama} {\bf 2020}, {\em 324},~1212--1213.

\bibitem[Epstude and Roese(2008)]{epstude2008functional}
Epstude, K.; Roese, N.J.
\newblock The functional theory of counterfactual thinking.
\newblock {\em Personality and social psychology review} {\bf 2008}, {\em
  12},~168--192.

\bibitem[Weidman(2014)]{weidman2014penetration}
Weidman, G.
\newblock {\em Penetration testing: a hands-on introduction to hacking}; No
  Starch Press,  2014.

\bibitem[Rajendran \em{et~al.}(2011)Rajendran, Jyothi, and
  Karri]{rajendran2011blue}
Rajendran, J.; Jyothi, V.; Karri, R.
\newblock Blue team red team approach to hardware trust assessment.
\newblock  2011 IEEE 29th international conference on computer design (ICCD).
  IEEE,  2011, pp. 285--288.

\bibitem[Rege(2016)]{rege2016incorporating}
Rege, A.
\newblock Incorporating the human element in anticipatory and dynamic cyber
  defense.
\newblock  2016 IEEE International Conference on Cybercrime and Computer
  Forensic (ICCCF). IEEE,  2016, pp. 1--7.

\bibitem[Ahmadpour \em{et~al.}(2019)Ahmadpour, Pedell, Mayasari, and
  Beh]{ahmadpour2019co}
Ahmadpour, N.; Pedell, S.; Mayasari, A.; Beh, J.
\newblock Co-creating and assessing future wellbeing technology using design
  fiction.
\newblock {\em She Ji: The Journal of Design, Economics, and Innovation} {\bf
  2019}, {\em 5},~209--230.

\bibitem[Pillai \em{et~al.}(2020)Pillai, Ahmadpour, Yoo, Kocaballi, Pedell,
  Sermuga~Pandian, and Suleri]{pillai2020communicate}
Pillai, A.G.; Ahmadpour, N.; Yoo, S.; Kocaballi, A.B.; Pedell, S.;
  Sermuga~Pandian, V.P.; Suleri, S.
\newblock {Communicate, Critique and Co-create (CCC) Future Technologies
  through Design Fictions in VR Environment}.
\newblock  Companion Publication of the 2020 ACM Designing Interactive Systems
  Conference,  2020, pp. 413--416.

\bibitem[Rapp(2020)]{rapp2020design}
Rapp, A.
\newblock {Design fictions for learning: A method for supporting students in
  reflecting on technology in Human-Computer Interaction courses}.
\newblock {\em Computers \& Education} {\bf 2020}, {\em 145},~103725.

\bibitem[Houde \em{et~al.}(2020)Houde, Liao, Martino, Muller, Piorkowski,
  Richards, Weisz, and Zhang]{houde2020business}
Houde, S.; Liao, V.; Martino, J.; Muller, M.; Piorkowski, D.; Richards, J.;
  Weisz, J.; Zhang, Y.
\newblock {Business (mis) Use Cases of Generative AI}.
\newblock {\em arXiv preprint arXiv:2003.07679} {\bf 2020}.

\bibitem[Carlini \em{et~al.}(2019)Carlini, Athalye, Papernot, Brendel, Rauber,
  Tsipras, Goodfellow, Madry, and Kurakin]{carlini2019evaluating}
Carlini, N.; Athalye, A.; Papernot, N.; Brendel, W.; Rauber, J.; Tsipras, D.;
  Goodfellow, I.; Madry, A.; Kurakin, A.
\newblock On evaluating adversarial robustness.
\newblock {\em arXiv preprint arXiv:1902.06705} {\bf 2019}.

\bibitem[John \em{et~al.}(2018)John, Glendenning, Marchant, Montgomery,
  Stewart, Wood, Lloyd, and Hawton]{john2018self}
John, A.; Glendenning, A.C.; Marchant, A.; Montgomery, P.; Stewart, A.; Wood,
  S.; Lloyd, K.; Hawton, K.
\newblock {Self-harm, suicidal behaviours, and cyberbullying in children and
  young people: Systematic review}.
\newblock {\em Journal of medical internet research} {\bf 2018}, {\em
  20},~e129.

\bibitem[Crothers(2020)]{crothers}
Crothers, B.
\newblock {FBI warns on teenage sextortion as new twists on sex-related scams
  emerge}.
\newblock
  \url{https://www.foxnews.com/tech/fbi-warns-teenage-sextortion-new-twists-sex-scams-emerge},
   2020.
\newblock {Fox News}; accessed 02-November-2020.

\bibitem[Nilsson \em{et~al.}(2019)Nilsson, Pepelasi, Ioannou, and
  Lester]{nilsson2019understanding}
Nilsson, M.G.; Pepelasi, K.T.; Ioannou, M.; Lester, D.
\newblock {Understanding the link between Sextortion and Suicide}.
\newblock {\em International Journal of Cyber Criminology} {\bf 2019}, {\em
  13},~55--69.

\bibitem[Haag and Salam(2017)]{haag}
Haag, M.; Salam, M.
\newblock {Gunman in ‘Pizzagate’ Shooting Is Sentenced to 4 Years in
  Prison}.
\newblock
  \url{https://www.nytimes.com/2017/06/22/us/pizzagate-attack-sentence.html},
  2017.
\newblock {The New York Times}; accessed 02-November-2017.

\bibitem[Bessi and Ferrara(2016)]{bessi2016social}
Bessi, A.; Ferrara, E.
\newblock {Social bots distort the 2016 US Presidential election online
  discussion}.
\newblock {\em First Monday} {\bf 2016}, {\em 21}.

\bibitem[Assenmacher \em{et~al.}(2020)Assenmacher, Clever, Frischlich, Quandt,
  Trautmann, and Grimme]{assenmacher2020demystifying}
Assenmacher, D.; Clever, L.; Frischlich, L.; Quandt, T.; Trautmann, H.; Grimme,
  C.
\newblock {Demystifying Social Bots: On the Intelligence of Automated Social
  Media Actors}.
\newblock {\em Social Media+ Society} {\bf 2020}, {\em 6},~2056305120939264.

\bibitem[Boneh \em{et~al.}(2019)Boneh, Grotto, McDaniel, and
  Papernot]{boneh2019relevant}
Boneh, D.; Grotto, A.J.; McDaniel, P.; Papernot, N.
\newblock {How relevant is the Turing test in the age of sophisbots?}
\newblock {\em IEEE Security \& Privacy} {\bf 2019}, {\em 17},~64--71.

\bibitem[Yang \em{et~al.}(2019)Yang, Varol, Davis, Ferrara, Flammini, and
  Menczer]{yang2019arming}
Yang, K.C.; Varol, O.; Davis, C.A.; Ferrara, E.; Flammini, A.; Menczer, F.
\newblock Arming the public with artificial intelligence to counter social
  bots.
\newblock {\em Human Behavior and Emerging Technologies} {\bf 2019}, {\em
  1},~48--61.

\bibitem[Shao \em{et~al.}(2018)Shao, Ciampaglia, Varol, Yang, Flammini, and
  Menczer]{shao2018spread}
Shao, C.; Ciampaglia, G.L.; Varol, O.; Yang, K.C.; Flammini, A.; Menczer, F.
\newblock The spread of low-credibility content by social bots.
\newblock {\em Nature communications} {\bf 2018}, {\em 9},~1--9.

\bibitem[Yan \em{et~al.}(2020)Yan, Yang, Menczer, and
  Shanahan]{yan2020asymmetrical}
Yan, H.Y.; Yang, K.C.; Menczer, F.; Shanahan, J.
\newblock Asymmetrical perceptions of partisan political bots.
\newblock {\em New Media \& Society} {\bf 2020}, p. 1461444820942744.

\bibitem[Farokhmanesh(2018)]{farokhmanesh2018legal}
Farokhmanesh, M.
\newblock {Is It Legal to Swap Someone’s Face into Porn without Consent?}
\newblock {\em Verge. January} {\bf 2018}, {\em 30}.

\bibitem[Karras \em{et~al.}(2020)Karras, Laine, Aittala, Hellsten, Lehtinen,
  and Aila]{karras2020analyzing}
Karras, T.; Laine, S.; Aittala, M.; Hellsten, J.; Lehtinen, J.; Aila, T.
\newblock {Analyzing and improving the image quality of StyleGAN}.
\newblock  Proceedings of the IEEE/CVF Conference on Computer Vision and
  Pattern Recognition,  2020, pp. 8110--8119.

\bibitem[Chen \em{et~al.}(2020)Chen, Yuan, Zhang, Zhao, Zhang, Chen, and
  Wang]{chen2020devil}
Chen, Y.; Yuan, X.; Zhang, J.; Zhao, Y.; Zhang, S.; Chen, K.; Wang, X.
\newblock {Devil’s whisper: A general approach for physical adversarial
  attacks against commercial black-box speech recognition devices}.
\newblock  29th USENIX Security Symposium (USENIX Security 20),  2020.

\bibitem[Duan \em{et~al.}(2020)Duan, Ma, Wang, Bailey, Qin, and
  Yang]{duan2020adversarial}
Duan, R.; Ma, X.; Wang, Y.; Bailey, J.; Qin, A.K.; Yang, Y.
\newblock {Adversarial Camouflage: Hiding Physical-World Attacks with Natural
  Styles}.
\newblock  Proceedings of the IEEE/CVF Conference on Computer Vision and
  Pattern Recognition,  2020, pp. 1000--1008.

\bibitem[Kong \em{et~al.}(2020)Kong, Guo, Li, and Liu]{kong2020physgan}
Kong, Z.; Guo, J.; Li, A.; Liu, C.
\newblock {PhysGAN: Generating Physical-World-Resilient Adversarial Examples
  for Autonomous Driving}.
\newblock  Proceedings of the IEEE/CVF Conference on Computer Vision and
  Pattern Recognition,  2020, pp. 14254--14263.

\bibitem[Nassi \em{et~al.}(2020)Nassi, Nassi, Ben-Netanel, Mirsky, Drokin, and
  Elovici]{nassi2020phantom}
Nassi, B.; Nassi, D.; Ben-Netanel, R.; Mirsky, Y.; Drokin, O.; Elovici, Y.
\newblock {Phantom of the ADAS: Phantom Attacks on Driver-Assistance Systems.}
\newblock {\em IACR Cryptol. ePrint Arch.} {\bf 2020}, {\em 2020},~85.

\bibitem[Wang \em{et~al.}(2020)Wang, Lv, Kuang, Zhao, Tan, Zhang, and
  Hu]{wang2020towards}
Wang, Y.; Lv, H.; Kuang, X.; Zhao, G.; Tan, Y.a.; Zhang, Q.; Hu, J.
\newblock {Towards a Physical-World Adversarial Patch for Blinding Object
  Detection Models}.
\newblock {\em Information Sciences} {\bf 2020}.

\bibitem[Rahman \em{et~al.}(2020)Rahman, Hossain, Alrajeh, and
  Alsolami]{rahman2020adversarial}
Rahman, A.; Hossain, M.S.; Alrajeh, N.A.; Alsolami, F.
\newblock {Adversarial examples--security threats to COVID-19 deep learning
  systems in medical IoT devices}.
\newblock {\em IEEE Internet of Things Journal} {\bf 2020}.

\bibitem[Ciosek(2020)]{ciosek2020aggravating}
Ciosek, I.
\newblock {AGGRAVATING UNCERTAINTY--RUSSIAN INFORMATION WARFARE IN THE WEST}.
\newblock {\em Torun International Studies} {\bf 2020}, {\em 1},~57--72.

\bibitem[Lamb(2020)]{lamb}
Lamb, A.
\newblock {After Covid, AI will Pivot}.
\newblock
  \url{https://towardsdatascience.com/after-covid-ai-will-pivot-dbe9dd06327},
  2020.
\newblock Towards data sciece; accessed 12-November-2020.

\bibitem[Smith and Rustagi(2020)]{vieve}
Smith, G.; Rustagi, I.
\newblock {The Problem With COVID-19 Artificial Intelligence Solutions and How
  to Fix Them}.
\newblock
  \url{https://ssir.org/articles/entry/the_problem_with_covid_19_artificial_intelligence_solutions_and_how_to_fix_them},
   2020.
\newblock Standford Social Innovation Review; accessed 12-November-2020.

\bibitem[Yampolskiy(2008)]{yampolskiy2008mimicry}
Yampolskiy, R.V.
\newblock Mimicry attack on strategy-based behavioral biometric.
\newblock  Fifth International Conference on Information Technology: New
  Generations (itng 2008). IEEE,  2008, pp. 916--921.

\bibitem[Yampolskiy and Govindaraju(2010)]{yampolskiy2010taxonomy}
Yampolskiy, R.V.; Govindaraju, V.
\newblock Taxonomy of behavioural biometrics. In {\em Behavioral Biometrics for
  Human Identification: Intelligent Applications}; IGI Global,  2010; pp.
  1--43.

\bibitem[Yampolskiy(2006)]{yampolskiy2006analyzing}
Yampolskiy, R.V.
\newblock Analyzing user password selection behavior for reduction of password
  space.
\newblock  Proceedings 40th Annual 2006 International Carnahan Conference on
  Security Technology. IEEE,  2006, pp. 109--115.

\bibitem[Whyte(2020)]{whyte2020deepfake}
Whyte, C.
\newblock {Deepfake news: AI-enabled disinformation as a multi-level public
  policy challenge}.
\newblock {\em Journal of Cyber Policy} {\bf 2020}, {\em 5},~199--217.

\bibitem[Goodfellow \em{et~al.}(2014)Goodfellow, Pouget-Abadie, Mirza, Xu,
  Warde-Farley, Ozair, Courville, and Bengio]{goodfellow2014generative}
Goodfellow, I.; Pouget-Abadie, J.; Mirza, M.; Xu, B.; Warde-Farley, D.; Ozair,
  S.; Courville, A.; Bengio, Y.
\newblock Generative adversarial nets.
\newblock  Advances in neural information processing systems,  2014, pp.
  2672--2680.

\bibitem[Tucciarelli \em{et~al.}(2020)Tucciarelli, Vehar, and
  Tsakiris]{tucciarelli2020realness}
Tucciarelli, R.; Vehar, N.; Tsakiris, M.
\newblock On the realness of people who do not exist: the social processing of
  artificial faces.
\newblock {\em PsyArXiv} {\bf 2020}.

\bibitem[Young(2019)]{young2019calibration}
Young, L.
\newblock {Calibration Camouflage: Hyphen-Labs and Adam Harvey: HyperFace}.
\newblock {\em Architectural Design} {\bf 2019}, {\em 89},~28--31.

\bibitem[Baggili and Behzadan(2019)]{baggili2019founding}
Baggili, I.; Behzadan, V.
\newblock {Founding The Domain of AI Forensics}.
\newblock {\em arXiv preprint arXiv:1912.06497} {\bf 2019}.

\bibitem[Schneider and Breitinger(2020)]{schneider2020ai}
Schneider, J.; Breitinger, F.
\newblock {AI Forensics: Did the Artificial Intelligence System Do It? Why?}
\newblock {\em arXiv preprint arXiv:2005.13635} {\bf 2020}.

\bibitem[Rosenberg \em{et~al.}(2013)Rosenberg, Halpern, Shulman, Wexler, and
  Phartiyal]{rosenberg2013reinvigorating}
Rosenberg, A.A.; Halpern, M.; Shulman, S.; Wexler, C.; Phartiyal, P.
\newblock Reinvigorating the role of science in democracy.
\newblock {\em PLoS Biol} {\bf 2013}, {\em 11},~e1001553.

\bibitem[{MIT Open Learning}(2020)]{moon}
{MIT Open Learning}.
\newblock {Tackling the misinformation epidemic with ``In Event of Moon
  Disaster" }.
\newblock
  \url{https://news.mit.edu/2020/mit-tackles-misinformation-in-event-of-moon-disaster-0720},
   2020.
\newblock {MIT News}; accessed 11-October-2020.

\bibitem[Fallis(2020)]{fallis2020epistemic}
Fallis, D.
\newblock {The Epistemic Threat of Deepfakes}.
\newblock {\em Philosophy \& Technology} {\bf 2020}, pp. 1--21.

\bibitem[Popper(2014)]{popper2014conjectures}
Popper, K.
\newblock {\em {Conjectures and refutations: The growth of scientific
  knowledge}}; routledge,  2014.

\bibitem[Deutsch(2011)]{deutsch2011beginning}
Deutsch, D.
\newblock {\em {The beginning of infinity: Explanations that transform the
  world}}; Penguin UK,  2011.

\bibitem[Baudrillard(1994)]{baudrillard1994simulacra}
Baudrillard, J.
\newblock {\em Simulacra and simulation}; University of Michigan press,  1994.

\bibitem[Hopf \em{et~al.}(2019)Hopf, Krief, Mehta, and Matlin]{hopf2019fake}
Hopf, H.; Krief, A.; Mehta, G.; Matlin, S.A.
\newblock Fake science and the knowledge crisis: ignorance can be fatal.
\newblock {\em Royal Society open science} {\bf 2019}, {\em 6},~190161.

\bibitem[D'Amour \em{et~al.}(2020)D'Amour, Heller, Moldovan, Adlam, Alipanahi,
  Beutel, Chen, Deaton, Eisenstein, Hoffman, et~al.]{d2020underspecification}
D'Amour, A.; Heller, K.; Moldovan, D.; Adlam, B.; Alipanahi, B.; Beutel, A.;
  Chen, C.; Deaton, J.; Eisenstein, J.; Hoffman, M.D.; others.
\newblock {Underspecification Presents Challenges for Credibility in Modern
  Machine Learning}.
\newblock {\em arXiv preprint arXiv:2011.03395} {\bf 2020}.

\bibitem[Oughton \em{et~al.}(2019)Oughton, Ralph, Pant, Leverett, Copic,
  Thacker, Dada, Ruffle, Tuveson, and Hall]{oughton2019stochastic}
Oughton, E.J.; Ralph, D.; Pant, R.; Leverett, E.; Copic, J.; Thacker, S.; Dada,
  R.; Ruffle, S.; Tuveson, M.; Hall, J.W.
\newblock {Stochastic Counterfactual Risk Analysis for the Vulnerability
  Assessment of Cyber-Physical Attacks on Electricity Distribution
  Infrastructure Networks}.
\newblock {\em Risk Analysis} {\bf 2019}, {\em 39},~2012--2031.

\bibitem[Almasoud \em{et~al.}(2020)Almasoud, Hussain, and
  Hussain]{almasoud2020smart}
Almasoud, A.S.; Hussain, F.K.; Hussain, O.K.
\newblock {Smart contracts for blockchain-based reputation systems: A
  systematic literature review}.
\newblock {\em Journal of Network and Computer Applications} {\bf 2020}, {\em
  170},~102814.

\bibitem[Cresci(2020)]{cresci2020decade}
Cresci, S.
\newblock A decade of social bot detection.
\newblock {\em Communications of the ACM} {\bf 2020}, {\em 63},~72--83.

\bibitem[Cresci \em{et~al.}(2017)Cresci, Di~Pietro, Petrocchi, Spognardi, and
  Tesconi]{cresci2017paradigm}
Cresci, S.; Di~Pietro, R.; Petrocchi, M.; Spognardi, A.; Tesconi, M.
\newblock {The paradigm-shift of social spambots: Evidence, theories, and tools
  for the arms race}.
\newblock  Proceedings of the 26th international conference on world wide web
  companion,  2017, pp. 963--972.

\bibitem[Barrett(2017)]{barrett2017theory}
Barrett, L.F.
\newblock The theory of constructed emotion: an active inference account of
  interoception and categorization.
\newblock {\em Social cognitive and affective neuroscience} {\bf 2017}, {\em
  12},~1--23.

\bibitem[Aliman(2020)]{alimansw2020}
Aliman, N.M.
\newblock {Self-Shielding Worlds}.
\newblock \url{https://nadishamarie.jimdo.com/clipboard/},  2020.
\newblock {Online}; accessed 23-November-2020.

\bibitem[TURING(1950)]{turing1950computing}
TURING, I.B.A.
\newblock {Computing machinery and intelligence-AM Turing}.
\newblock {\em Mind} {\bf 1950}, {\em 59},~433.

\bibitem[Pantserev(2020)]{pantserev2020malicious}
Pantserev, K.A.
\newblock {The Malicious Use of AI-Based Deepfake Technology as the New Threat
  to Psychological Security and Political Stability}. In {\em Cyber Defence in
  the Age of AI, Smart Societies and Augmented Humanity}; Springer,  2020; pp.
  37--55.

\bibitem[{\"O}hman(2019)]{ohman2019introducing}
{\"O}hman, C.
\newblock {Introducing the pervert’s dilemma: a contribution to the critique
  of Deepfake Pornography}.
\newblock {\em Ethics and Information Technology} {\bf 2019}, pp. 1--8.

\bibitem[Macaulay(2020)]{hybri}
Macaulay, T.
\newblock {New AR app will let you model a virtual companion on anyone you
  want}.
\newblock
  \url{https://thenextweb.com/neural/2020/06/01/new-ar-app-will-let-you-model-a-virtual-companion-on-anyone-you-want/},
   2020.
\newblock Online; accessed 04-August-2020.

\bibitem[Kumar \em{et~al.}(2020)Kumar, Nystr{\"o}m, Lambert, Marshall,
  Goertzel, Comissoneru, Swann, and Xia]{kumar2020adversarial}
Kumar, R.S.S.; Nystr{\"o}m, M.; Lambert, J.; Marshall, A.; Goertzel, M.;
  Comissoneru, A.; Swann, M.; Xia, S.
\newblock {Adversarial Machine Learning--Industry Perspectives}.
\newblock {\em arXiv preprint arXiv:2002.05646} {\bf 2020}.

\bibitem[Barrett and Simmons(2015)]{barrett2015interoceptive}
Barrett, L.F.; Simmons, W.K.
\newblock Interoceptive predictions in the brain.
\newblock {\em Nature reviews neuroscience} {\bf 2015}, {\em 16},~419--429.

\bibitem[Kleckner \em{et~al.}(2017)Kleckner, Zhang, Touroutoglou, Chanes, Xia,
  Simmons, Quigley, Dickerson, and Barrett]{kleckner2017evidence}
Kleckner, I.R.; Zhang, J.; Touroutoglou, A.; Chanes, L.; Xia, C.; Simmons,
  W.K.; Quigley, K.S.; Dickerson, B.C.; Barrett, L.F.
\newblock Evidence for a large-scale brain system supporting allostasis and
  interoception in humans.
\newblock {\em Nature human behaviour} {\bf 2017}, {\em 1},~1--14.

\bibitem[Aliman and Kester(2019)]{aliman2019requisite}
Aliman, N.; Kester, L.
\newblock {Requisite Variety in Ethical Utility Functions for {AI} Value
  Alignment}.
\newblock  Proceedings of the Workshop on Artificial Intelligence Safety 2019
  co-located with the 28th International Joint Conference on Artificial
  Intelligence, AISafety@IJCAI 2019, Macao, China, August 11-12, 2019.,  2019.

\bibitem[Dignum(2020)]{dignum2020ai}
Dignum, V.
\newblock {AI is multidisciplinary}.
\newblock {\em AI Matters} {\bf 2020}, {\em 5},~18--21.

\bibitem[Floridi(2019)]{floridi2019establishing}
Floridi, L.
\newblock {Establishing the rules for building trustworthy AI}.
\newblock {\em Nature Machine Intelligence} {\bf 2019}, {\em 1},~261--262.

\bibitem[Hagendorff(2020)]{hagendorff2020ethics}
Hagendorff, T.
\newblock {The ethics of Ai ethics: An evaluation of guidelines}.
\newblock {\em Minds and Machines} {\bf 2020}, pp. 1--22.

\bibitem[Mittelstadt(2019)]{mittelstadt2019ai}
Mittelstadt, B.
\newblock {AI Ethics--Too principled to fail}.
\newblock {\em arXiv preprint arXiv:1906.06668} {\bf 2019}.

\bibitem[Whittlestone \em{et~al.}(2019)Whittlestone, Nyrup, Alexandrova, and
  Cave]{whittlestone2019role}
Whittlestone, J.; Nyrup, R.; Alexandrova, A.; Cave, S.
\newblock {The role and limits of principles in AI ethics: towards a focus on
  tensions}.
\newblock  Proceedings of the 2019 AAAI/ACM Conference on AI, Ethics, and
  Society,  2019, pp. 195--200.

\bibitem[Jobin \em{et~al.}(2019)Jobin, Ienca, and Vayena]{jobin2019global}
Jobin, A.; Ienca, M.; Vayena, E.
\newblock {The global landscape of AI ethics guidelines}.
\newblock {\em Nature Machine Intelligence} {\bf 2019}, {\em 1},~389--399.

\bibitem[Gu \em{et~al.}(2014)Gu, Konana, Raghunathan, and Chen]{gu2014research}
Gu, B.; Konana, P.; Raghunathan, R.; Chen, H.M.
\newblock {Research note—The allure of homophily in social media: Evidence
  from investor responses on virtual communities}.
\newblock {\em Information Systems Research} {\bf 2014}, {\em 25},~604--617.

\bibitem[Yoo(2007)]{yoo2007ideological}
Yoo, J.
\newblock {Ideological Homophily and Echo Chamber Effect in Internet and Social
  Media}.
\newblock {\em Student International Journal of Research} {\bf 2007}, {\em
  4},~1--7.

\bibitem[Tsao \em{et~al.}(2019)Tsao, Ting, and Johnson]{tsao2019creative}
Tsao, J.; Ting, C.; Johnson, C.
\newblock Creative outcome as implausible utility.
\newblock {\em Review of General Psychology} {\bf 2019}, {\em 23},~279--292.

\bibitem[Yampolskiy \em{et~al.}(2012)Yampolskiy, Ashby, and
  Hassan]{yampolskiy2012wisdom}
Yampolskiy, R.V.; Ashby, L.; Hassan, L.
\newblock {Wisdom of Artificial Crowds—A Metaheuristic Algorithm for
  Optimization}.
\newblock {\em Journal of Intelligent Learning Systems and Applications} {\bf
  2012}, {\em 4},~98--107.

\bibitem[Yampolskiy(2020)]{guideyam}
Yampolskiy, R.
\newblock {Usable Guidelines Aim to Make AI Safer}.
\newblock
  \url{{https://www.mouser.com/blog/usable-guidelines-aim-to-make-ai-safer}},
  2020.
\newblock {All, EIT 2020: The Intelligent Revolution}; accessed
  13-November-2020.

\bibitem[Trazzi and Yampolskiy(2018)]{trazzi2018building}
Trazzi, M.; Yampolskiy, R.V.
\newblock {Building safer AGI by introducing artificial stupidity}.
\newblock {\em arXiv preprint arXiv:1808.03644} {\bf 2018}.

\bibitem[Trazzi and Yampolskiy(2020)]{trazzi2020artificial}
Trazzi, M.; Yampolskiy, R.V.
\newblock {Artificial Stupidity: Data We Need to Make Machines Our Equals}.
\newblock {\em Patterns} {\bf 2020}, {\em 1},~100021.

\bibitem[Aliman and Kester(2020)]{aliman2020artificial}
Aliman, N.M.; Kester, L.
\newblock Artificial creativity augmentation.
\newblock  International Conference on Artificial General Intelligence.
  Springer,  2020, pp. 23--33.

\bibitem[Leviathan and Matias(2018)]{49194}
Leviathan, Y.; Matias, Y.
\newblock {Google Duplex: An AI System for Accomplishing Real-World Tasks Over
  the Phone}.
\newblock
  \url{https://ai.googleblog.com/2018/05/duplex-ai-system-for-natural-conversation.html},
   2018.

\bibitem[Yampolskiy(2020)]{yampolskiy2020controllability}
Yampolskiy, R.V.
\newblock {On Controllability of AI}.
\newblock {\em arXiv preprint arXiv:2008.04071} {\bf 2020}.

\bibitem[Yampolskiy(2019)]{yampolskiy2019unpredictability}
Yampolskiy, R.V.
\newblock {Unpredictability of AI}.
\newblock {\em arXiv preprint arXiv:1905.13053} {\bf 2019}.

\bibitem[Bostr{\"o}m(2014)]{bostrom2014superintelligence}
Bostr{\"o}m, N.
\newblock {Superintelligence: Paths, dangers, strategies}.
\newblock {\em Oxford University Press} {\bf 2014}.

\bibitem[Baum \em{et~al.}(2017)Baum, Barrett, and Yampolskiy]{baum2017modeling}
Baum, S.; Barrett, A.; Yampolskiy, R.V.
\newblock Modeling and interpreting expert disagreement about artificial
  superintelligence.
\newblock {\em Informatica} {\bf 2017}, {\em 41},~419--428.

\bibitem[Friston(2018)]{friston2018self}
Friston, K.
\newblock {Am I self-conscious?(Or does self-organization entail
  self-consciousness?)}.
\newblock {\em Frontiers in psychology} {\bf 2018}, {\em 9},~579.

\bibitem[Bruineberg \em{et~al.}(2018)Bruineberg, Kiverstein, and
  Rietveld]{bruineberg2018anticipating}
Bruineberg, J.; Kiverstein, J.; Rietveld, E.
\newblock The anticipating brain is not a scientist: the free-energy principle
  from an ecological-enactive perspective.
\newblock {\em Synthese} {\bf 2018}, {\em 195},~2417--2444.

\bibitem[Rudrauf \em{et~al.}(2017)Rudrauf, Bennequin, Granic, Landini, Friston,
  and Williford]{rudrauf2017mathematical}
Rudrauf, D.; Bennequin, D.; Granic, I.; Landini, G.; Friston, K.; Williford, K.
\newblock A mathematical model of embodied consciousness.
\newblock {\em Journal of theoretical biology} {\bf 2017}, {\em 428},~106--131.

\bibitem[Williford \em{et~al.}(2018)Williford, Bennequin, Friston, and
  Rudrauf]{williford2018projective}
Williford, K.; Bennequin, D.; Friston, K.; Rudrauf, D.
\newblock The projective consciousness model and phenomenal selfhood.
\newblock {\em Frontiers in Psychology} {\bf 2018}, {\em 9},~2571.

\bibitem[Deutsch(2013)]{deutsch2013constructor}
Deutsch, D.
\newblock Constructor theory.
\newblock {\em Synthese} {\bf 2013}, {\em 190},~4331--4359.

\bibitem[Deutsch and Marletto(2015)]{deutsch2015constructor}
Deutsch, D.; Marletto, C.
\newblock Constructor theory of information.
\newblock {\em Proceedings of the Royal Society A: Mathematical, Physical and
  Engineering Sciences} {\bf 2015}, {\em 471},~20140540.

\bibitem[Dietrich(2015)]{dietrich2015creativity}
Dietrich, A.
\newblock {\em How creativity happens in the brain}; Springer,  2015.

\bibitem[Brockman(2020)]{brockman2020possible}
Brockman, J.
\newblock {\em Possible minds: Twenty-five ways of looking at AI. Beyond Reward
  and Punishment. David Deutsch.}; Penguin Books,  2020.

\bibitem[Reynolds and Lewis(2017)]{reynolds2017teams}
Reynolds, A.; Lewis, D.
\newblock Teams solve problems faster when they’re more cognitively diverse.
\newblock {\em Harvard Business Review} {\bf 2017}, {\em 30}.

\bibitem[Friston \em{et~al.}(2017)Friston, Lin, Frith, Pezzulo, Hobson, and
  Ondobaka]{friston2017active}
Friston, K.J.; Lin, M.; Frith, C.D.; Pezzulo, G.; Hobson, J.A.; Ondobaka, S.
\newblock Active inference, curiosity and insight.
\newblock {\em Neural computation} {\bf 2017}, {\em 29},~2633--2683.

\bibitem[Sajid \em{et~al.}(2019)Sajid, Ball, and Friston]{sajid2019active}
Sajid, N.; Ball, P.J.; Friston, K.J.
\newblock Active inference: demystified and compared.
\newblock {\em arXiv} {\bf 2019}, pp. arXiv--1909.

\bibitem[Hernandez \em{et~al.}(2020)Hernandez, Marin-Castro,
  et~al.]{hernandez2020semantic}
Hernandez, J.; Marin-Castro, H.M.; others.
\newblock {A Semantic Focused Web Crawler Based on a Knowledge Representation
  Schema}.
\newblock {\em Applied Sciences} {\bf 2020}, {\em 10},~3837.

\bibitem[Singh \em{et~al.}(2020)Singh, Tomar, and Sangaiah]{singh2020sentiment}
Singh, N.K.; Tomar, D.S.; Sangaiah, A.K.
\newblock Sentiment analysis: a review and comparative analysis over social
  media.
\newblock {\em Journal of Ambient Intelligence and Humanized Computing} {\bf
  2020}, {\em 11},~97--117.

\bibitem[Kowsari \em{et~al.}(2019)Kowsari, Jafari~Meimandi, Heidarysafa, Mendu,
  Barnes, and Brown]{kowsari2019text}
Kowsari, K.; Jafari~Meimandi, K.; Heidarysafa, M.; Mendu, S.; Barnes, L.;
  Brown, D.
\newblock {Text classification algorithms: A survey}.
\newblock {\em Information} {\bf 2019}, {\em 10},~150.

\bibitem[Aggarwal \em{et~al.}(2019)Aggarwal, Woolley, Chabris, and
  Malone]{aggarwal2019impact}
Aggarwal, I.; Woolley, A.W.; Chabris, C.F.; Malone, T.W.
\newblock The impact of cognitive style diversity on implicit learning in
  teams.
\newblock {\em Frontiers in psychology} {\bf 2019}, {\em 10},~112.

\bibitem[den Houting(2019)]{den2019neurodiversity}
den Houting, J.
\newblock Neurodiversity: An insider’s perspective,  2019.

\bibitem[Blume(1998)]{blume1998neurodiversity}
Blume, H.
\newblock {Neurodiversity: On the neurological underpinnings of geekdom}.
\newblock {\em The Atlantic} {\bf 1998}, {\em 30}.

\bibitem[Chapman(2020)]{chapman2020neurodiversity}
Chapman, R.
\newblock Neurodiversity, disability, wellbeing.
\newblock {\em Neurodiversity Studies: A New Critical Paradigm} {\bf 2020}.

\bibitem[Chen \em{et~al.}(2019)Chen, Liu, Zhang, and Kwan]{chen2019cognitive}
Chen, X.; Liu, J.; Zhang, H.; Kwan, H.K.
\newblock {Cognitive diversity and innovative work behaviour: The mediating
  roles of task reflexivity and relationship conflict and the moderating role
  of perceived support}.
\newblock {\em Journal of Occupational and Organizational Psychology} {\bf
  2019}, {\em 92},~671--694.

\bibitem[Bolis \em{et~al.}(2017)Bolis, Balsters, Wenderoth, Becchio, and
  Schilbach]{bolis2017beyond}
Bolis, D.; Balsters, J.; Wenderoth, N.; Becchio, C.; Schilbach, L.
\newblock Beyond autism: introducing the dialectical misattunement hypothesis
  and a bayesian account of intersubjectivity.
\newblock {\em Psychopathology} {\bf 2017}, {\em 50},~355--372.

\bibitem[Abu-Akel \em{et~al.}(2020)Abu-Akel, Webb, de~Montpellier,
  Von~Bentivegni, Luechinger, Ishii, and Mohr]{abu2020autistic}
Abu-Akel, A.; Webb, M.E.; de~Montpellier, E.; Von~Bentivegni, S.; Luechinger,
  L.; Ishii, A.; Mohr, C.
\newblock Autistic and positive schizotypal traits respectively predict better
  convergent and divergent thinking performance.
\newblock {\em Thinking Skills and Creativity} {\bf 2020}, p. 100656.

\bibitem[Paola \em{et~al.}(2020)Paola, Laura, Giusy, and
  Michela]{paola2020autism}
Paola, P.; Laura, G.; Giusy, M.; Michela, C.
\newblock Autism, autistic traits and creativity: a systematic review and
  meta-analysis.
\newblock {\em Cognitive Processing} {\bf 2020}, pp. 1--36.

\bibitem[Kasirer \em{et~al.}(2020)Kasirer, Adi-Japha, and
  Mashal]{10.3389/fpsyg.2020.559238}
Kasirer, A.; Adi-Japha, E.; Mashal, N.
\newblock {Verbal and Figural Creativity in Children With Autism Spectrum
  Disorder and Typical Development}.
\newblock {\em Frontiers in Psychology} {\bf 2020}, {\em 11},~2968.
\newblock
  doi:{\changeurlcolor{black}\href{https://doi.org/10.3389/fpsyg.2020.559238}{\detokenize{10.3389/fpsyg.2020.559238}}}.

\bibitem[Hoogman \em{et~al.}(2020)Hoogman, Stolte, Baas, and
  Kroesbergen]{hoogman2020creativity}
Hoogman, M.; Stolte, M.; Baas, M.; Kroesbergen, E.
\newblock {Creativity and ADHD: A review of behavioral studies, the effect of
  psychostimulants and neural underpinnings}.
\newblock {\em Neuroscience \& Biobehavioral Reviews} {\bf 2020}.

\bibitem[White(2020)]{white2020thinking}
White, H.A.
\newblock {Thinking “Outside the Box”: Unconstrained Creative Generation in
  Adults with Attention Deficit Hyperactivity Disorder}.
\newblock {\em The Journal of Creative Behavior} {\bf 2020}, {\em
  54},~472--483.

\bibitem[White and Shah(2016)]{white2016scope}
White, H.A.; Shah, P.
\newblock {Scope of semantic activation and innovative thinking in college
  students with ADHD}.
\newblock {\em Creativity Research Journal} {\bf 2016}, {\em 28},~275--282.

\bibitem[Greenberg and Arndt(2011)]{greenberg2011terror}
Greenberg, J.; Arndt, J.
\newblock Terror management theory.
\newblock {\em Handbook of theories of social psychology} {\bf 2011}, {\em
  1},~398--415.

\bibitem[Solomon \em{et~al.}(2015)Solomon, Greenberg, and
  Pyszczynski]{solomon2015worm}
Solomon, S.; Greenberg, J.; Pyszczynski, T.
\newblock {\em {The worm at the core: On the role of death in life}}; Random
  House,  2015.

\bibitem[Chittaro \em{et~al.}(2017)Chittaro, Sioni, Crescentini, and
  Fabbro]{chittaro2017mortality}
Chittaro, L.; Sioni, R.; Crescentini, C.; Fabbro, F.
\newblock Mortality salience in virtual reality experiences and its effects on
  users’ attitudes towards risk.
\newblock {\em International Journal of Human-Computer Studies} {\bf 2017},
  {\em 101},~10--22.

\bibitem[Shehryar and Hunt(2005)]{shehryar2005terror}
Shehryar, O.; Hunt, D.M.
\newblock A terror management perspective on the persuasiveness of fear
  appeals.
\newblock {\em Journal of consumer psychology} {\bf 2005}, {\em 15},~275--287.

\end{thebibliography}




\end{document}